\documentclass[final]{siamltex}
\usepackage{showlabels}
\usepackage{xcolor}
\usepackage{amsmath}
\usepackage{amssymb}
\usepackage{mathrsfs}
\usepackage{graphicx}
 \usepackage{bm}
\usepackage{color}
\usepackage{float}
\usepackage{epstopdf}
\usepackage{algorithm,algorithmic}
\usepackage[breaklinks,colorlinks=true,linkcolor=blue,citecolor=red,backref=page]{hyperref}

\DeclareMathAlphabet{\itbf}{OML}{cmm}{b}{it}
 \DeclareMathAlphabet\mathbfcal{OMS}{cmsy}{b}{n}

\renewcommand{\hat}{\widehat}
\renewcommand{\tilde}{\widetilde}
\def\RR{\mathbb{R}}

\def\bx{{{\itbf x}}}

\def\by{{{\itbf y}}}
\def\bz{{{\itbf z}}}

\def\bxi{{\boldsymbol{\xi}}}
\def\om{{\omega}}

\def\bh{{\itbf{h}}}

\def\bH{{\itbf{H}}}
\def\cK{\mathcal{K}}
\def\hpar{h_{\parallel}}
\def\Hpar{H_{\parallel}}

\def\bK{{\boldsymbol{\kappa}}}

\def\xpar{x_{\parallel}}
\def\zpar{z_{\parallel}}
\def\ypar{y_{\parallel}}
\def\kapar{\kappa_{\parallel}}
\def\lb{\left <}
\def\rb{\right >}
\def\cI{\mathcal{I}}
\def\cS{\mathcal{S}}
\def\mI{\mathbb{K}}
\def\cJ{\mathcal{J}}
\def\cX{\mathcal{X}}
\def\cT{\mathcal{T}}
\def\cC{\mathcal{C}}
\def\cF{\mathcal{F}}

\def\cK{\mathcal{K}}
\def\cL{\mathcal{L}}
\def\cD{\mathcal{D}}

\def\CI{{\protect\scalebox{0.5}[0.4]{CI}}}
\def\SP{{\protect\scalebox{0.5}[0.4]{SP}}}
\def\PR{{\protect\scalebox{0.5}[0.4]{PR}}}
\def\OP{{\protect\scalebox{0.5}[0.4]{OP}}}
\def\CINT{{\protect\scalebox{0.5}[0.4]{CINT}}}

\def\SAR{{\protect\scalebox{0.5}[0.4]{SAR}}}

\def\EE{\mathbb{E}}
\def\la{\lambda}

\begin{document}

\title{Imaging in random media by two-point coherent interferometry} \author{Liliana Borcea\footnotemark[1] \and Josselin
  Garnier\footnotemark[2]} 
  
\maketitle


\renewcommand{\thefootnote}{\fnsymbol{footnote}}
\footnotetext[1]{Department of Mathematics, University of Michigan,
  Ann Arbor, MI 48109. {\tt borcea@umich.edu}}
\footnotetext[2]{CMAP, CNRS, Ecole Polytechnique, Institut Polytechnique de Paris, 91128 Palaiseau Cedex, France.  {\tt
    josselin.garnier@polytechnique.edu}}
\markboth{L. BORCEA, J. GARNIER}{Imaging in random media}

\begin{abstract}
This paper considers wave-based imaging through a heterogeneous  (random) scattering medium.
The goal is to  estimate the support of the reflectivity function of a remote scene from measurements of the backscattered wave field. The proposed imaging methodology is based on the coherent interferometric  (CINT) approach that exploits the local empirical cross correlations of the measurements of the wave field.  The standard CINT images are known to be robust (statistically stable) with respect to the random medium, but the stability comes at the expense of a loss of resolution. 
This paper shows that a two-point CINT function contains the information needed to obtain
statistically stable and high-resolution  images. Different methods to build such images are presented, theoretically analyzed and compared with the standard imaging approaches using numerical simulations.
The first method involves a phase-retrieval step to extract the reflectivity function from the modulus of its Fourier transform.
The second method involves the evaluation of the leading eigenvector of the two-point CINT imaging function seen as the kernel of a linear operator. The third method uses an optimization step to extract the reflectivity function from some cross products of its Fourier transform. The presentation is for the synthetic aperture radar data acquisition setup, where a moving sensor probes the scene with signals emitted periodically and records the resulting backscattered wave. The generalization to other imaging setups, with passive or active arrays of sensors, is discussed briefly. 
\end{abstract}

\section{Introduction}
\label{sec:Intro}
The goal of inverse scattering is to estimate an unknown medium using  sensors that emit 
probing  signals and measure the generated (backscattered or transmitted) waves. The mathematical formulation is an inverse problem for a wave equation, that seeks to ``invert" the nonlinear mapping between the unknown coefficients (e.g. wave speed) and the solution (the wave) at the points of measurement. Typically the mapping is not invertible, hence the quotation marks, and only an approximation of the coefficients can be expected. 

Wave-based imaging is a simplification of the inverse scattering 
problem, which seeks a qualitative estimate of the reflective  structures in the medium, modeled by 
a reflectivity function $\rho$ which captures sharp variations (e.g., jump discontinuities)  of the coefficients in the wave equation. It uses various data processing to compute an imaging function (image in short) $\cI(\by)$ at points $\by$ in a search region, whose peaks are interpreted as  locations of the sought-after  reflectors.

The existing imaging methodology has been driven by  applications in radar imaging \cite{curlander1991synthetic,cheney2009fundamentals,gilman2017transionospheric}, seismic inversion \cite{claerbout1985imaging,biondi20063d}, nondestructive evaluation \cite{shull2002nondestructive}, 
ultrasound for medical diagnostics \cite{jensen2007medical}, etc. It consists of  efficient methods 
known as reverse time migration \cite{biondi20063d}, filtered backprojection \cite{cheney2009fundamentals}, linear sampling 
\cite{cakoni2005qualitative}, factorization \cite{kirsch2008factorization} and so on. These methods assume that the background medium which hosts the reflective structures  is known and non-scattering. With a few exceptions, they also  use the single-scattering (Born) approximation, which linearizes the mapping from the reflectivity $\rho$ to the wave at the sensors. The Born approximation 
is only partially understood \cite[Section 3]{symes2009seismic} but it is the foundation of contemporary seismic and radar imaging. Multiple scattering effects  are known to cause some artifacts aka ``ghosts" or ``multiples"  in images,  but these are sometimes easy to spot and they pale in comparison with the  artifacts caused by heterogeneous backgrounds. 

We are interested in imaging through heterogeneous media with numerous weak inhomogeneities, 
like concrete, biological tissue, the Earth's crust and the turbulent atmosphere.  The inhomogeneities, which are often called ``clutter",  have little effect on waves at low enough frequency and at moderate imaging range, so the existing  methodology is adequate.  For example, S-band radar  which operates at 2-4GHz is mostly insensitive to atmospheric effects.
However, the pursuit of better resolution of images, which is proportional to the wavelength,  has lead to the development of high-frequency systems like Ka-band radar (26.5-40GHz)  and even W-band radar (75-110GHz). Cumulative 
scattering  in the atmosphere impedes imaging with such high-frequency radar, especially  at ranges $\ge 1$km of interest.
Conventional approaches ignore this impediement and thus produce images with significant and unpredictable artifacts. 

To study imaging through heterogeneous backgrounds,  we model the inhomogeneities by random fluctuations of the wave speed. Imaging is carried out in a single medium, not an ensemble of realizations,  but the inhomogeneities are impossible to estimate from the band-limited data and are therefore uncertain. The random model places the problem in a stochastic framework, where we can quantify the robustness of images to the uncertainty  of the fluctuations of the wave speed.  

As long as the range of imaging does not exceed a transport mean free path, the length scale marking 
the onset of diffusion in the random medium \cite{van1999multiple}, coherent imaging can be done with empirical cross correlations of the measurements calculated over carefully chosen pairs of sensors (less than  $\cX$ apart) and time windows of duration $\cT$. The coherent interferometric (CINT) imaging approach  \cite{borcea2006adaptive,borcea2011enhanced,garnier2008coherent}
synchronizes these cross correlations relative to an imaging point $\by$, using travel time delays calculated in the known reference medium, without fluctuations, and then sums over the selected pairs of sensors to compute the image $\cI^{\CINT}(\by)$. The user-defined thresholds $\cX$ and $\cT$ are key in CINT, as they account for the statistical decorrelation of the components of the wave field \cite{rytov1989principles,van1999multiple} over a spatial offset $\cX_d$ called the decoherence length, 
and a frequency offset  $\Omega_d$ called the decoherence frequency. These scales depend on the statistics of the fluctuations of the wave speed, not the particular realization, and since the optimal threshold choices are $\cX \lesssim \cX_d$ and $1/\cT \lesssim \Omega_d$, they can be estimated  with optimization,  while forming CINT images  \cite{borcea2006adaptive}. The resolution analysis of CINT is  well understood \cite{borcea2006adaptive,borcea2011enhanced} and  reveals a trade-off between resolution and robustness, also called statistical stability with respect to the uncertainty of the medium. The robustness is controlled by the user-defined  $\cX$ and $1/\cT$:  the smaller they are, the less sensitive the image $\cI^{\CINT}(\by)$ to the inhomogeneities in the medium. However,  the diameter of the support of the CINT point spread function  is roughly proportional  to $1/\cX$ in the cross-range direction and $\cT$ in the range direction, so robustness comes at the expense of resolution. Worse yet, the resolution cannot be  improved by increasing the aperture that supports the sensors and the bandwidth of the probing signals. 

It was observed in \cite{borcea2018passive}, in the context of imaging a group of point sources with a passive array
of receivers, that CINT does not use the empirical cross correlations in an optimal way. A different processing of these cross correlations led to a new ``two-point CINT" function $\cI(\by,\by')$, which allowed accurate estimates of offsets between the source locations.  Imaging was then carried out with a direct search algorithm,  but the search  grows in complexity with the 
number of sources, and the localization is correct up to a rigid body translation and rotation of the group.

A similar two-point CINT function $\cI(\by,\by')$ was introduced and analyzed in \cite{borcea2020high}  for imaging with synthetic aperture radar (SAR) in random media. The main result was that it is possible to estimate from it the modulus $|\hat \rho(\bK)|$ of the Fourier transform $\hat \rho(\bK)$ of the reflectivity $\rho(\by)$, for $\bK$ in a rectangular domain centered at the origin, with side lengths determined by the central wavelength and the bandwidth of the probing signals, the aperture size and  the range of imaging. Phase retrieval was then used to estimate $\hat \rho(\bK)$, followed by an inverse Fourier transform to 
get an image. This approach is robust to both cumulative scattering and additive noise induced effects, as long as there is strong prior
information about the reflectivity, such as $\rho(\by) \ge 0$. Without it, the method is unstable and prone to noise artifacts. 
In addition, the estimation is correct only up to a rigid body translation and a reflection with 
respect to the origin.

In  this paper we 
show
 that the two-point CINT function $\cI(\by,\by')$ contains more information about the unknown reflectivity than has  been used so far. 
 To exploit this information we introduce two new imaging approaches: 

\vspace{0.03in}
1. The first approach is specialized to an imaging scene with  a group of small, point-like reflectors. 
It uses the  spectral decomposition of the linear integral operator $\cL$ with kernel  $\cI(\by,\by')$,  and shows that if the reflectors are not too close, then they can be localized by using the leading eigenvector of $\cL$ with better resolution than in CINT. This resolution is not as good as the one achieved in \cite{borcea2020high} with the phase retrieval approach, but there are important advantages: The method is much more robust to clutter and noise effects, there is no restriction on the sign of the reflectivity, and there is no translation or reflection ambiguity. 
In particular, unlike any
of the existing methods, it can distinguish between reflectors with positive and negative reflectivities.

\vspace{0.03in}
2. The second approach applies to a general reflectivity function $\rho(\by)$. It estimates from the data 
the product $\hat \rho(\bK) \overline{\hat \rho(\bK')}$, where the bar denotes complex conjugate and  i) the vector 
$(\bK+\bK')/2$ lies in a rectangular domain centered at the origin, of side lengths determined by the central wavelength and the bandwidth of the probing signals, the aperture size and  the range of imaging, ii) the vector $\bK-\bK'$ lies in a much smaller  rectangular domain, also centered at the origin, with side lengths that depend on the decoherence parameters  $\cX_d$, $\Omega_d$ and the user-defined thresholds $\cX$, $\cT$. 
The Fourier transform of the reflectivity can then be estimated from these products using optimization, and an image follows via inverse Fourier transform. 

\vspace{0.03in}
We study these two imaging approaches from first principles, using analysis and numerical simulations, in the context of SAR imaging in random media. We also explain briefly how the results extend to imaging  sources with passive arrays of receivers and also to imaging reflectivities with active arrays of sensors that are both sources and receivers. The random wave distortion caused by scattering in the heterogeneous background is captured in our analysis
using the high-frequency, geometrical optics model of wave propagation in random media \cite{rytov1989principles}. This model has been used for imaging  in \cite{borcea2011enhanced,garnier2016passive, borcea2018passive,borcea2020high},
so we recall the relevant results from there. The advantage of the model is that it accounts for the  main  impediments of
coherent imaging in random media, but it is simple enough to allow a very explicit calculation of the two-point CINT function $\cI(\by,\by')$. Nevertheless, our imaging approaches are not model specific, and they can be analyzed with more complex 
wave propagation models, like the paraxial one studied in \cite{garnier2016fourth}.

The paper is organized as follows: Section \ref{sect:Form} formulates the SAR imaging problem and reviews from \cite{borcea2020high}  the calculation and properties of the two-point CINT function $\cI(\by,\by')$. The new results are in sections \ref{sect:spectral}--\ref{sect:num}. The first imaging approach is studied in section \ref{sect:spectral} and 
the second imaging approach is given in section 
\ref{sect:Wigner}. We use numerical simulations in section \ref{sect:num} to assess the performance of these two approaches and compare with conventional imaging, CINT imaging and imaging using phase retrieval. We end with a summary in section \ref{sect:sum}.

\section{Formulation of the imaging problem}
\label{sect:Form}
We give here the SAR data model for imaging, and recall from \cite{borcea2020high} the relevant facts on the two-point CINT function, that are used in sections \ref{sect:spectral}--\ref{sect:Wigner}.  We also describe briefly how the results generalize to two other common imaging setups with passive and active arrays of sensors.

\subsection{SAR imaging in random media}
\label{sect:SAR}
In SAR imaging, a sensor placed on a moving platform emits periodically, at time interval $T$, 
a signal {$f(t)$} which generates a wave that propagates to the reflective imaging scene, is backscattered there and is then recorded back at the sensor. The location of the sensor at the $n^{\rm th}$ emission instant $nT$ 
is denoted by $\bx_n$, and the recording is $R_n(t)$, for time $t \in (0,T)$. The trajectory of the platform may be arbitrary, but for simplicity we assume it is straight, of length $a$, called the aperture.  We make two hypotheses that allow for two simplifying approximations:\\
1) The platform moves slowly so that the start-stop approximation is valid. This considers the platform as stationary during the round trip of the wave from the sensor to the reflective imaging scene, and it is justified by the large wave speed compared to the speed of the platform. However, the motion of the sensor between the emission and reception can be accounted for, by introducing some  Doppler phase corrections in the analysis, as is done for example in \cite{borcea2017resolution}.\\
2) The platform moves a small distance 
in each period $T$, so the locations $\bx_n$ of emission and recording are close enough
to allow the  approximation of sums over $n = 0, \ldots, N$ by integrals over the aperture. This so-called continuum 
aperture approximation is convenient for the analysis. 

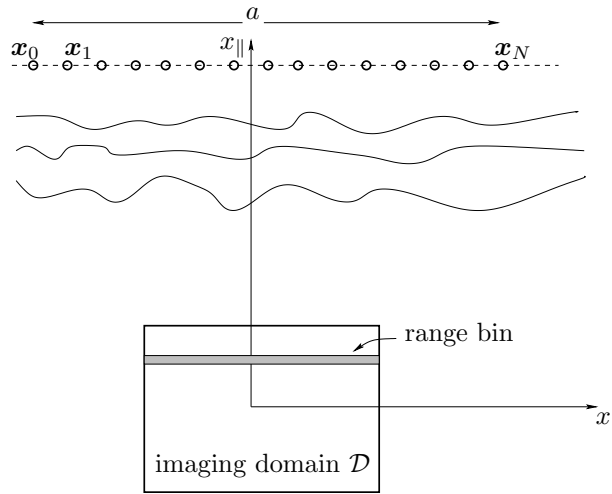
\begin{figure}[t]
\vspace{-3.5in}
\begin{picture}(0,0)%
\hspace{1.5in}\includegraphics[width = 0.6\textwidth]{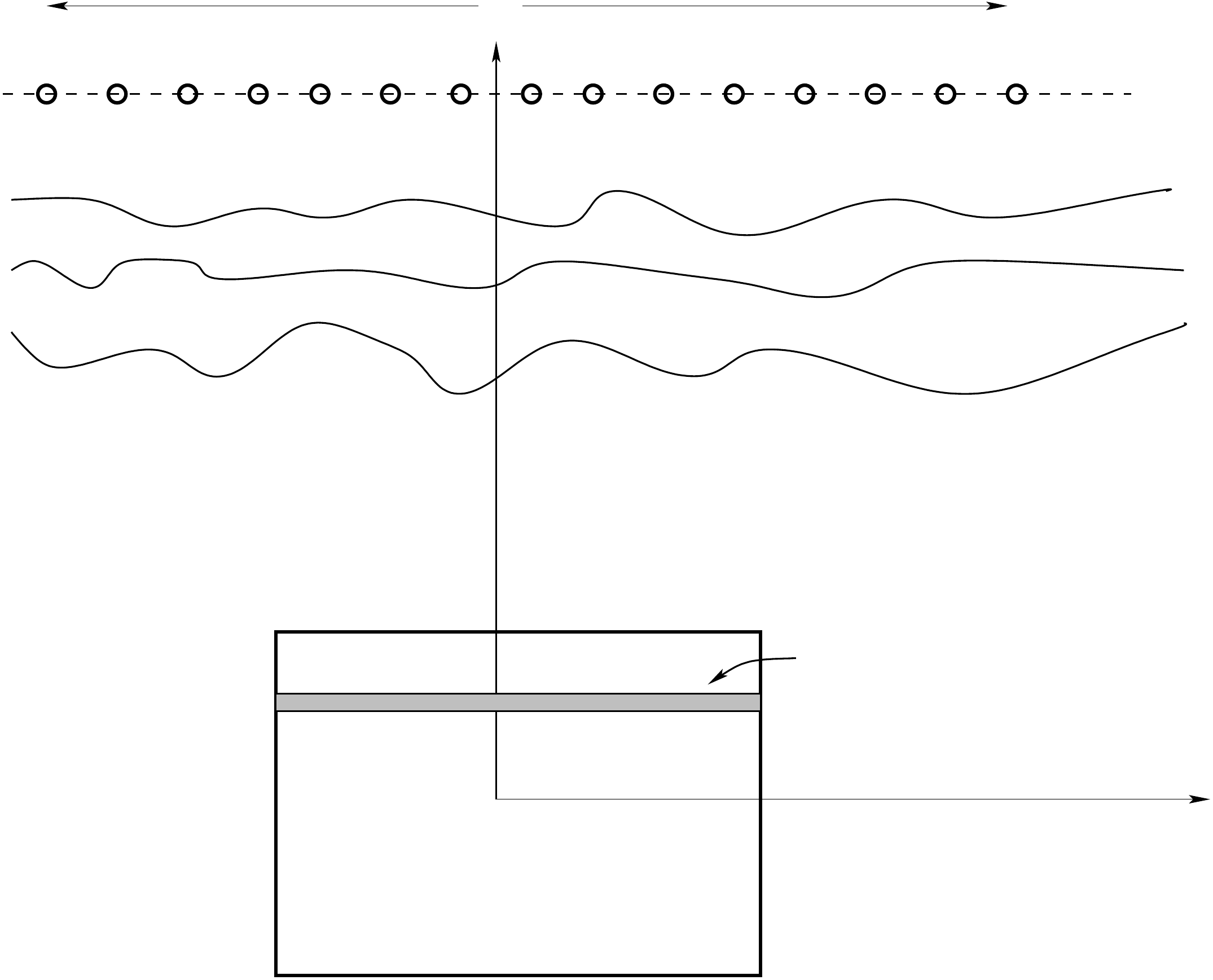}%
\end{picture}%
\setlength{\unitlength}{2960sp}%
\begingroup\makeatletter\ifx\SetFigFont\undefined%
\gdef\SetFigFont#1#2#3#4#5{%
  \reset@font\fontsize{#1}{#2pt}%
  \fontfamily{#3}\fontseries{#4}\fontshape{#5}%
  \selectfont}%
\fi\endgroup%
\begin{picture}(10799,8069)(1599,-7583)
\put(4000,-3900){\makebox(0,0)[lb]{\smash{{\SetFigFont{7}{8.4}{\familydefault}{\mddefault}{\updefault}{\color[rgb]{0,0,0}{\normalsize $\bx_0$}}%
}}}}
\put(4450,-3900){\makebox(0,0)[lb]{\smash{{\SetFigFont{7}{8.4}{\familydefault}{\mddefault}{\updefault}{\color[rgb]{0,0,0}{\normalsize $\bx_1$}}%
}}}}
\put(8050,-3900){\makebox(0,0)[lb]{\smash{{\SetFigFont{7}{8.4}{\familydefault}{\mddefault}{\updefault}{\color[rgb]{0,0,0}{\normalsize $\bx_N$}}%
}}}}
\put(8900,-7000){\makebox(0,0)[lb]{\smash{{\SetFigFont{7}{8.4}{\familydefault}{\mddefault}{\updefault}{\color[rgb]{0,0,0}{\normalsize $x$}}%
}}}}
\put(5750,-3850){\makebox(0,0)[lb]{\smash{{\SetFigFont{7}{8.4}{\familydefault}{\mddefault}{\updefault}{\color[rgb]{0,0,0}{\normalsize $x_\parallel$}}%
}}}}
\put(5970,-3600){\makebox(0,0)[lb]{\smash{{\SetFigFont{7}{8.4}{\familydefault}{\mddefault}{\updefault}{\color[rgb]{0,0,0}{\normalsize $a$}}%
}}}}
\put(5210,-7400){\makebox(0,0)[lb]{\smash{{\SetFigFont{7}{8.4}{\familydefault}{\mddefault}{\updefault}{\color[rgb]{0,0,0}{\normalsize imaging domain $\cD$}}%
}}}}
\put(7300,-6300){\makebox(0,0)[lb]{\smash{{\SetFigFont{7}{8.4}{\familydefault}{\mddefault}{\updefault}{\color[rgb]{0,0,0}{\normalsize range bin}}%
}}}}

\end{picture}%
\vspace{0.1in}
\caption{Illustration of the SAR imaging setup in two dimensions.  The sensor 
moves on a straight trajectory which defines a linear aperture of length $a$. It emits signals and records the backscattered waves at points $\bx_j$, for $j = 0, \ldots, N$. 
The imaging domain supports 
the unknown reflectivity and it is separated from the aperture by a random medium. The  hashed region  in the imaging domain $\cD$ illustrates a range bin. The drawing is not up to scale. The diameter of $\cD$ is supposed to be much smaller than the distance (range) to the sensor, 
which is why the range bin is approximately a thin horizontal strip. }
\label{fig:setup}
\end{figure}

The signal {$f(t)$} may be 
a chirp \cite{curlander1991synthetic} or a pulse, with bandwidth $B$ and modulation at  central frequency $\om_o \gg B$.  We use a pulse and assume for convenience that it has Gaussian envelope
\begin{equation}
{f(t) = s(t) +c.c.} , \quad\quad
s(t) =  \frac{B}{\sqrt{2 \pi}} \exp \Big(-i \om_o t-\frac{B^2 t^2}{2} \Big)  ,
\label{eq:F1} 
\end{equation}
where $c.c.$ stands for complex conjugate.
All the recordings are gathered in the data used for imaging:
\begin{equation}
{\rm data} = \{  
R_n(t), 
~~ t \in (0,T), ~~ n=0,\ldots, N\}.
\label{eq:F.2}
\end{equation}
Note that there are only two degrees of freedom in these data: the ``slow time" $nT$ of emission of the 
signals and the ``fast time" $t$. Thus, it is impossible to determine a reflectivity function in three dimensions,
which is why imaging is done on a surface with known topography \cite{curlander1991synthetic,cheney2009fundamentals}.
For simplicity, we  work in two dimensions, where  the problem is formally determined\footnote{In three dimensions, the two-dimensional image in the range vs. cross-range plane can be mapped  to an image on the surface with known topography.}.
We use the  system of coordinates $\bx = (x,\xpar)$ with origin at the center of the  imaging domain $\cD$, as illustrated in Fig. \ref{fig:setup}. The $\xpar$ axis is along the ``range" direction that points from $\cD$ to the aperture and the $x$ axis is along the ``cross-range" direction, which is aligned with the aperture. 

SAR data are often processed using range gating, which  groups the recordings based on their arrival times.
It allows the segmentation of $\cD$  in ``range bins", which are sets of points with range estimated from the selected arrival time  window (see Fig. \ref{fig:setup}). The advantage of range gating is that imaging in each range bin is one-dimensional, in the cross-range direction. We will use it in both the analysis and the numerical simulations.

\subsection{Wave equation}
\label{sect:Form.1}
The wave generated by the $n^{\rm th}$ emission is modeled by 
{$u_{{\rm real},n}(\bx,t)=u_n(\bx,t)+c.c.$, where $u_n(\bx,t)$ is the solution of the wave equation}
\begin{equation}
\frac{1}{c^2} \left[ 1 + \sigma \mu \left(\frac{\bx}{\ell}\right) + \rho(\bx) \right] \partial_t^2 u_n(\bx,t) - \Delta_{\bx} u_n(\bx,t) = 
s(t-nT) \delta(\bx-\bx_n), 
\label{eq:F3} 
\end{equation}
for $t \in \RR$ and $\bx = (x,\xpar) \in \RR^2$.
The wave field is vanishing before the onset of the source. 
The reflectivity that we wish to image is  $\rho(\bx)$ and the wave speed in the heterogeneous background medium 
fluctuates about the constant  value $c$. The fluctuations are modeled by the random field
$\mu(\bxi)$ of dimensionless argument $\bxi \in \RR^2$. This field is statistically homogeneous, with 
mean zero and integrable covariance 
\begin{equation}
\cC(\bxi-\bxi') = \EE [\mu(\bxi) \mu(\bxi') ],
\label{eq:F5}
\end{equation}
and it is normalized so that 
\begin{equation}
\cC({\bf 0}) =1 ~~ \mbox{and}~~ 
\int_{\RR^2} d \bxi \, \cC(\bxi) = 1.
\label{eq:F6}
\end{equation}
The length scale $\ell$ in \eqref{eq:F3} is the correlation length (typical size of the inhomogeneities) and $\sigma$ is the standard deviation of the fluctuations.
For the sake of explicit calculations, we use the Gaussian covariance function
\begin{equation}
\cC(\bxi\, ) = \exp(-\pi |\bxi|^2).
\label{eq:F7}
\end{equation}
Other covariance functions can be considered, like the Kolmogorov model \cite[Appendix 4.A.4]{gilman2017transionospheric},  and the imaging will be qualitatively the same.

The wave propagation model  is based on the high-frequency, long-range scale separation assumption 
\begin{equation}
\la_o \ll \ell \ll L,
\label{eq:F8}
\end{equation}
where $\la_o = 2 \pi c/\om_o$ is the central wavelength and $L$ is the range scale i.e., the distance from the aperture to 
the range bin in the imaging domain. The standard deviation of the fluctuations is small $\sigma \ll 1$, but still causes large random wavefront distortion. More details
on the scaling assumptions are given in Appendix \ref{ap:A}, where we recall from \cite{garnier2016passive} the 
random geometrical optics approximation of the Green's function used  next  to write the data model.

\subsection{Data model}
\label{sect:Form.2}
We model the data  
\begin{align}\label{eq:F9a}
R_{n}(t) &= u_{{\rm real},n}(\bx_n,t) + W_n(t) , \qquad u_{{\rm real},n}(\bx_n,t) 
=u_n(\bx_n,t)+c.c.
\end{align}
using the single-scattering (Born) approximation with respect to the reflectivity $\rho$, as is typically done in SAR imaging:
\begin{align}
u_n(\bx_n,t)  & \approx \int_{\RR} \frac{d \om}{2 \pi} e^{-i \om(t-nT)} \hat s(\om) k^2(\om) \int_{\cD} d \bz \, 
 \rho(\bz) \hat G_\mu^2(\bz,\bx_n,\om) .
 \label{eq:F9}
 \end{align}
Here $\hat s(\om)$ is the Fourier transform of the signal \eqref{eq:F1}
$$
\hat s(\om) = \exp\Big( - \frac{(\om-\om_o)^2}{2B^2}\Big) ,
$$
$k(\om) = \om/c$ is the wavenumber, the propagation in the random medium from 
$\bx_n$ to  $\bz \in \mbox{supp}(\rho) \subset \cD$ is modeled by the Green's function $\hat G_\mu$ of the Helmholtz's equation with squared index of refraction $1 + \sigma \mu(\bx/\ell)$, and $W_n(t)$ is additive measurement noise. 

The random travel time model gives (see Appendix \ref{ap:A})
\begin{equation}
\hat G_\mu(\bz,\bx_n,\om) \approx \hat G_o(\bz,\bx_n,\om) \exp \left[ i \om \tau_\mu(\bz,\bx_n) \right],
\label{eq:A2}
\end{equation}
where 
\begin{equation}
\hat G_o(\bz,\bx_n,\om) = \frac{i}{4} H_0^{(1)} \big(k(\om) |\bz-\bx_n|\big) \approx \frac{\exp \big[ i k(\om) |\bz-\bx_n| + i \pi/4 \big]}{
\sqrt{8 \pi k(\om) |\bz-\bx_n|}},
\label{eq:A3}
\end{equation}
is the Green's function in the reference homogeneous medium with wave speed $c$ and $\tau_\mu(\bz,\bx_n)$ is the random travel time fluctuation which gives the wavefront distortion.
As explained in Appendix \ref{ap:A}, due to the long-range assumption $|\bz-\bx_n| = O(L) \gg \ell$, 
the process $\tau_\mu(\bz,\bx_n)$ has Gaussian statistics, so we can easily calculate the statistical moments 
of $\hat G_\mu^2$. We write here the first two moments: The mean is negligible
\begin{equation}
\EE \big[ \hat G_\mu^2 (\bz,\bx_n,\om) \big] \approx 0,
\label{eq:Mom1}
\end{equation}
due to the large random wavefront distortion. This behavior is 
well known and it is called the loss of coherence of the wave due to scattering in the random medium \cite{rytov1989principles,ishimaru2017electromagnetic}.
The second moment, where the bar denotes complex conjugate, is 
\begin{align}
\EE \big[ \hat G_\mu^2 (\bz,\bx_n,\om) \overline{\hat G_\mu^2 (\bz',\bx_{n'},\om')}  \big] \approx &
\hat G_o^2 (\bz,\bx_n,\om) \overline{\hat G_o^2 (\bz',\bx_{n'},\om')} \nonumber \\
&
\times \exp \Big[ - \frac{(\om-\om')^2 }{2\Omega_d^2} - \frac{ (x_{n'}-x_n)^2 }{2 \cX_d^2} \Big].
\label{eq:Mom2}
\end{align}
Unlike the mean, this moment is large as long as the frequency and sensor offsets 
are smaller than the  decoherence frequency $\Omega_d$ and decoherence length $\cX_d$, respectively. 
These two scales model the decorrelation of the wave components in the random medium. They are written explicitly in Appendix \ref{ap:A} and satisfy\footnote{The summary in Appendix \ref{ap:A} shows that $\Omega_d \ll \om_o$ and $\cX_d \ll \ell$. 
For broadband imaging systems it is typical that $\Omega_d \ll B$. We choose the aperture 
$a \lesssim \ell$ in order to simplify the expression of the imaging functions. In practice this can be arranged 
by aperture segmentation, and then combining the results obtained with each sub-aperture.}  
\begin{equation}
\Omega_d \ll B \ll \om_o, ~~ \cX_d \ll  a \lesssim \ell.
\label{eq:DecScales}
\end{equation}
Note that \eqref{eq:Mom2} is a simplification of a more general second moment formula given in 
Appendix \ref{ap:A}, with right-hand side depending on $\bz$ and $\bz'$. This simplification holds 
if  the cross-range diameter of the support of $\rho$, which is a subset of 
the imaging domain $\cD$, is smaller than $\cX_d$. We refer to the end of Appendix \ref{ap:A} for a justification
of  this assumption  in the context of this paper.

The additive noise $W_n(t)$ is modeled as Gaussian,  mean zero, white in $t$ and uncorrelated in $n$. Its Fourier 
components $\hat W_n(\om)$ satisfy 
\begin{equation}
\EE \left[ \hat W_n(\om) \overline{ \hat W_{n'}(\om')} \right] = \sigma_{_W}^2 \delta(\om-\om') \delta_{n,n'},
\label{eq:Mom3}
\end{equation}
where $\delta_{n,n'}$ is the Kronecker symbol.  Other additive noise models can be considered, 
and the results will be qualitatively similar.

\subsection{The two-point CINT function}
\label{sect:2CINT}
Let $\by \in \cD$ be a point in the imaging domain and introduce the following reference wave field 
\begin{equation}
F_n(\by,t) = \int_{\RR} \frac{d \om}{2 \pi} \, e^{-i \om t} \hat F_n(\by,\om), \qquad \hat F_n(\by,\om):= \hat s(\om) \hat G_o^2 (\by,\bx_n,\om) 
\exp\Big(- \frac{x_n^2}{a^2}\Big).
\label{eq:Mom4}
\end{equation}
Aside from the exponential factor, which is a Gaussian apodization convenient for the analysis, this field 
models the wave emitted and received at $\bx_n$, after scattering by a hypothetic point reflector at $\by$ in the reference medium.   
In conventional imaging, $F_n(\by,t)$ is ``matched" to the data in order to obtain the 
SAR imaging function
\begin{align}
\cI^{\SAR}(\by)&:= \sum_{n=0}^N \int_\RR dt \, R_{n}(t) \overline{F_n(\by,t-nT)} \nonumber \\
&= \sum_{n=0}^N \int_\RR \frac{d \om}{2 \pi} \hat R_n(\om) \overline{\hat F_n(\by,\om)} e^{- i \om n T}.
\label{eq:ISAR}
\end{align}

The two-point CINT function is calculated as follows \cite{borcea2020high}
\begin{align}
\cI(\by,\by') :=  \sum_{n,n'=0}^N \int_\RR &\frac{d \om}{2 \pi} \int_\RR \frac{d \om'}{2 \pi}
 \hat R_n(\om) \overline{\hat F_n(\by,\om)} e^{- i \om n T}  \overline{ \hat R_{n'}(\om')} 
 \hat F_{n'}(\by',\om') e^{ i \om' n' T} \nonumber \\
& \times \exp \left[ -\frac{(x_n-x_{n'})^2}{2 \cX^2} - \frac{(\om-\om')^2}{2 \Omega^2} \right], \qquad \by,\by' \in \cD,
 \label{eq:2CINT}
\end{align}
where $\cX$ and $\Omega = 1/\cT$ are the user-defined threshold parameters discussed in the introduction. 
Because it depends on two points, \eqref{eq:2CINT} is not an imaging function, so we need 
additional processing to obtain an image. 
The simplest  processing is to evaluate \eqref{eq:2CINT}
on the diagonal, and thus obtain the CINT image
\begin{equation}
\cI^{\CINT}(\by) :=  \cI(\by,\by) .
\label{eq:CINT}
\end{equation}
However, this is not the optimal choice, as we explain  in the next section, where we study the expression of the expectation of \eqref{eq:2CINT}. The derivation of this expression is a lengthy calculation given in \cite{borcea2020high}, which uses the continuum aperture approximation.

Note that $\cI^{\CINT}(\by)$ is nonnegative, real valued (see Appendix \ref{app:proofICINT}) 
and furthermore, $\cI^{\CINT}(\by) = \left|\cI^{\SAR}(\by)\right|^2$ if we choose $\cX  \gg a, ~\Omega \gg B$.
We will see that the thresholding is important
for achieving statistical stability of  the two-point CINT  function, meaning that its random fluctuations are small with respect 
to its mean. A good choice of the threshold parameters turns out to be  $\cX \lesssim \cX_d$ and $\Omega \lesssim \Omega_d$, so that the expectations
of all the terms in \eqref{eq:2CINT} are large as  we will discuss in section \ref{sect:stab}.

\subsection{The expectation of the two-point CINT function}

The calculations in \cite{borcea2020high} give that 
\begin{equation} 
\EE \left[ \cI(\by,\by') \right] \approx \cJ(\by,\by')e^{-2 i k_o(\ypar-\ypar')} + \EE\left[\cI_{_W}(\by,\by')\right],
\label{eq:2.1}
\end{equation}
where 
\begin{align}
\cJ(\by,\by') = C \int_{\cD} d \bz \int_{\cD} d \bz' \rho_{k_o}(\bz) \overline{\rho_{k_o}(\bz')}\, \cK_{\bH}\Big( \frac{\bz+\bz'}{2} 
-\frac{\by+\by'}{2} \Big) \nonumber \\
\times  \cK_{\bh}\left( (\bz-\bz')-(\by-\by') \right),
\label{eq:2.2}
\end{align}
and the second term is the contribution of the additive noise
\begin{align}
\EE\left[\cI_{_W}(\by,\by')\right] = C_{_W} \exp \left[-\frac{(\ypar-\ypar')^2}{\hpar^2} - \frac{(y-y')^2}{2 h^2} - 2 i k_o(\ypar-\ypar')\right].
\label{eq:2.3}
\end{align}
The calculation is carried out with the continuum approximation of a dense array 
$\sum_{n=0}^N \psi(x_n) \exp(-x_n^2/a^2)  
\approx \int_\RR dx \,\psi(x) \exp(-x^2/a^2)$.
In \eqref{eq:2.2} we introduced the reflectivity modulated in the range direction at the central wavenumber $k_o = \om_o/c$,
\begin{equation}
\rho_{k_o}(\bz):= \rho(\bz) \exp(-2 i k_o \zpar),
\label{eq:2.4}
\end{equation}
and  the blurring kernels 
\begin{align}
\cK_{\bH}(\bx) = K_H(x) K_{\Hpar}(\xpar), \qquad \cK_{\bh}(\bx) = K_h(x) K_{\hpar}(\xpar),
\label{eq:KernH}
\end{align}
defined by the Gaussian
\begin{equation}
K_\alpha(x) := \frac{1}{\sqrt{2 \pi} \alpha} \exp\Big(-\frac{x^2}{2 \alpha^2} \Big), \qquad \alpha \in \{H,\Hpar, h,\hpar\}.
\label{eq:kernels}
\end{equation}
The resolution scales in \eqref{eq:KernH} are 
\begin{align}
\bH &= (H,\Hpar), \quad H:= \frac{L}{2 k_o} \sqrt{\frac{1}{\cX^2} + \frac{1}{\cX_d^2} + \frac{1}{a^2}}, \quad 
\Hpar := \frac{c}{2} \sqrt{\frac{1}{\Omega^2} + \frac{1}{\Omega_d^2} + \frac{1}{B^2}}, \label{eq:resH} 
\end{align}
and 
\begin{align}
\bh &= (h,\hpar), \quad h:= \frac{L}{k_o a}, \quad 
\hpar := \frac{c}{B}. \label{eq:resh}
\end{align}
The factors $C$ and $C_{_W}$ in \eqref{eq:2.2}--\eqref{eq:2.3} are constants whose expressions are given in \cite{borcea2020high} and whose precise values are not important for this paper. We only 
need the ratio 
\begin{equation}
\frac{C_{_W}}{C} = O(\sigma_{_W}^2),
\label{eq:relW}
\end{equation}
which quantifies the relative strength of the additive noise effect. 

Note that  if imaging took place in the reference medium 
($\Omega_d\to+\infty$, ${\cal X}_d\to +\infty$) and we had no thresholding ($\Omega\to+\infty$, ${\cal X}\to +\infty$), then the resolution scales would satisfy
$\bH = \bh/2$, and we would get after a straightforward calculation that 
\begin{align}
\cJ_o(\by,\by')&= C_o \left[\int_\cD d \bz \,  \rho_{k_o}(\bz) \cK_{\bh/\sqrt{2}}(\bz-\by) \right]  \overline{\left[\int_\cD d \bz' \, \rho_{k_o}(\bz') \cK_{\bh/\sqrt{2}}(\bz'-\by') \right]} \nonumber \\
&= \cI^{\SAR}_o(\by) \overline{\cI^{\SAR}_o(\by')}
.
\label{eq:HOMMed}
\end{align}
Here $C_o$ is a constant similar to $C$ and $\cI^{\SAR}_o(\by) $ is the ideal SAR image in the homogeneous reference medium, with well-known resolution described by $\bh=(h,\hpar)$  \cite{curlander1991synthetic,gilman2017transionospheric}. Moreover, when evaluating \eqref{eq:HOMMed} at $\by = \by'$, we  recover the squared modulus of the conventional SAR imaging function.

Cumulative scattering in the random medium impedes imaging, and this is captured in 
\eqref{eq:2.2} by the kernel 
$\cK_{\bH}$, with resolution scales \eqref{eq:resH} satisfying 
\begin{align}
\frac{H}{h} &= \frac{1}{2} \sqrt{1 + \Big(\frac{a}{\cX}\Big)^2 + \Big(\frac{a}{\cX_d} \Big)^2} \stackrel{\eqref{eq:DecScales}}{\gg} 1, \label{eq:2.6}\\
\frac{\Hpar}{\hpar} &= \frac{1}{2} \sqrt{1 + \Big(\frac{B}{\Omega}\Big)^2 + \Big(\frac{B}{\Omega_d}\Big)^2} \stackrel{\eqref{eq:DecScales}}{\gg} 1. \label{eq:2.7}
\end{align}
 Therefore, equation \eqref{eq:2.2}
 shows that  for any 
$\bz, \bz' \in \mbox{supp}(\rho)$, the two-point CINT function can determine the center points $(\bz+\bz')/2$ with much less precision than the offsets $\bz-\bz'$.  

When evaluating \eqref{eq:2.1} on the diagonal, as we do in CINT imaging, we get 
\begin{equation}
\EE\left[ \cI^{\CINT}(\by) \right] \approx C \int_{\cD} d \bz \int_{\cD} d \bz' \, \rho_{k_o}(\bz) 
\overline{\rho_{k_o}(\bz')} \cK_{\bh}(\bz-\bz') \cK_{\bH}\Big( \frac{\bz+\bz'}{2} 
-\by \Big) +C_{_W}.
\label{eq:ECINT}
\end{equation}
The information on the offsets $\bz-\bz'$ is lost in CINT, and the image is  
blurry, with the poor resolution  $\bH = (H,\Hpar)$. The point of \cite{borcea2018passive,borcea2020high} and this paper is that the  offset information 
contained in the two-point CINT function can be used to improve the resolution.

\subsection{Statistical stability}
\label{sect:stab}
The advantage of using the thresholding in the two-point CINT function is seen when studying 
the variance $\mbox{Var}[\cI(\by,\by')]$. This is calculated in  
\cite{borcea2020high} and we now summarize the results.

If there is no additive noise and if $\cX \lesssim \cX_d$, $\Omega \lesssim \Omega_d$, then we have
\begin{equation}
\sqrt{ \mbox{Var}[\cI(\by,\by')] } = \sqrt{O \left(\frac{\cX^2}{\cX_d^2}\right) + 
O \left(\frac{\Omega^2}{\Omega_d^2} \right)} 
\big| \EE [\cI(\by,\by')] \big|,
\label{eq:Var2CINT}
\end{equation}
so the smaller $\cX$ and $\Omega$ are, the closer $\cI(\by,\by')$ is to its mean. However, 
if  these are too small, the resolution scales in $\bH$ deteriorate. The optimal choice of 
the thresholds, which balances the  trade-off between resolution and stability, is, therefore, $\cX \lesssim \cX_d$ and $\Omega \lesssim \Omega_d$. In the analysis carried out in sections \ref{sect:spectral}--\ref{sect:Wigner} we assume  optimal thresholds, and approximate 
the two-point CINT function by its mean.

Note that in conventional SAR imaging there is no thresholding and the coefficient of variation of $\left|\cI^{\SAR}(\by)\right|^2$ is equal to one when  there is no additive noise \cite{borcea2020high}.
 The numerical simulations in section \ref{sect:num} illustrate how the statistical instability of conventional SAR manifests: the images change dramatically with the realizations of the fluctuations $\mu$, they have large artifacts and thus cannot be used to estimate the support of the reflectivity. 

For the additive noise contribution we give the covariance of the fluctuations 
\begin{align}
\mbox{Cov} \left[\cI_{_W}(\by,\by'), \cI_{_W}(\bz,\bz') \right] &= C_{_W}^2 
\exp \Big[ -\frac{(\ypar-\zpar)^2}{\hpar^2} - \frac{(y-z)^2}{2 h^2} + 2 i k_o(\ypar-\zpar)\Big] \nonumber \\
& \hspace{-0.3in} \times \exp \Big[ -\frac{(\ypar'-\zpar')^2}{\hpar^2} - \frac{(y'-z')^2}{2 h^2} - 2 i k_o(\ypar'-\zpar')\Big] ,
\label{eq:CovW}
\end{align}
which simplifies  for the CINT imaging function to 
\begin{align}
\mbox{Cov} \left[\cI^\CINT_{_W}(\by), \cI^\CINT_{_W}(\bz) \right] &= C_{_W}^2 
\exp \Big[ -\frac{2(\ypar-\zpar)^2}{\hpar^2} - \frac{(y-z)^2}{h^2} \Big]. 
\label{eq:CovWCINT}
\end{align}
Since $\left|\EE\left[\cI_{_W}(\by,\by')\right]\right|^2 = O(C_{_W}^2)$,  the noise contribution to two-point or to regular CINT imaging is not statistically stable. It causes speckle 
with relative amplitude $C_{_W}/C = O(\sigma_{_W}^2)$ and  
with speckle size of order $h$ in cross range and $\hpar$ in range. Such speckle is not important for the CINT image, 
which has the poor resolution quantified by $\bH$. In fact, the speckle can be removed by using 
a low-pass filter on $\cI^\CINT(\by)$.  However, noise affects the two-point CINT function $\cI(\by,\by')$, because the 
speckle size is similar to the resolution $\bh$ of  the offset vectors $\by-\by'$. If the noise is 
weak ($\sigma_{_W} \ll 1$),  it can be viewed as a perturbation of the imaging methods described in \cite{borcea2020high} and in section \ref{sect:Wigner}, which predict the ideal resolution quantified by $\bh$. However, if $\sigma_{_W}$ is larger, the speckle should be filtered out using a 
low-pass filter.  We can  still achieve a better resolution than with CINT, but not as good as the ideal one. 

\subsection{Generalization to other imaging setups}
\label{sect:2CINTgen}
The imaging methods described in sections \ref{sect:spectral}--\ref{sect:Wigner} are based 
on the expression \eqref{eq:2.2}, which gives the mean \eqref{eq:2.1} of the two-point CINT function.
We now describe how this expression generalizes to two other imaging setups:

\vspace{0.03in} $\bullet$ \textbf{Imaging sources with a passive array.}
Suppose that we had  a collection of $N+1$ receivers at closely spaced locations $\big(\bx_j\big)_{j=0}^N$ in a line segment of length $a$. Such a collection is called a passive array with aperture $a$.  The imaging problem would then be
to localize a group of sources in the domain $\cD$, which gives a source term in the wave equation of the form $S(\bx,t)$ that generates a wave {$u(\bx,t)$}.
This wave propagates through the random medium and it is received at the array. The data model 
\begin{align*}
R_n(t) :&= {u(\bx_n,t) }+ W_n(t)  , 
\end{align*}
with 
\begin{align*}
{u(\bx_n,t) }&= \int_{\RR} \frac{d \om}{2 \pi} e^{-i \om t}  \int_{\cD} d \bz \,\hat{S}(\bz, \om) 
 \hat G_\mu(\bz,\bx_n,\om) , \label{eq:G2}
 \end{align*}
is very similar to \eqref{eq:F9}, except for the one way propagation between the sources  and the receivers, which is why in the equation above we have $\hat G_\mu$ and not $\hat G_\mu^2$.  The two-point CINT function $\cI(\by,\by')$ is given by the same 
data processing as in \eqref{eq:2CINT}, with 
\begin{equation*}
\hat{F}_n(\by,\omega) =  \hat s(\om) \hat G_o(\by,\bx_n,\om) 
\exp\Big( -\frac{x_n^2}{a^2}\Big),
\end{equation*}
and {$\hat s(\om) = \exp[-(\om-\om_o)^2/(2B^2)] $} a frequency filter, with $B \ll \om_o$. The expression of the expectation 
of  $\cI(\by,\by')$ is \eqref{eq:2.1} modified as follows: $\rho(\bz)$ is replaced by $\hat{S}(\bz,\om_o)$ (assuming $\hat S(\bz,\om)$ is almost constant over the frequency interval  centered at $\om_o$ and bandwidth $B$), and 
all the resolution parameters in \eqref{eq:resH}--\eqref{eq:resh} are divided by two.

\vspace{0.03in} $\bullet$ \textbf{Imaging with an active array.}
If we wished to image the reflectivity $\rho$ using an active array of sensors at $\bx_n$, for $n = 0, \ldots, N$, 
 with linear aperture of length $a$, 
the data would be collected as follows: The array would emit a probing signal 
{$f(t)=s(t)+c.c.$} from one sensor location $\bx_n$ at a time, and the response 
\begin{equation*}
R_{n',n}(t) = u_{{\rm real}}(\bx_{n'} ,t; \bx_n) +W_{n',n}(t), 
\end{equation*}
with 
$$
u_{{\rm real}}(\bx_{n'} ,t; \bx_n) = \int_{\RR} \frac{d \om}{2 \pi} e^{-i \om t} \hat s(\om) k^2(\om) \int_{\cD} d \bz \, \rho(\bz) 
\hat G_\mu(\bz,\bx_n,\om) \hat G_\mu(\bz,\bx_{n'},\om) +c.c.
$$
would be measured at all the sensor locations $\bx_{n'}$, for $n' = 0, \ldots, N$, where $W_{n',n}(t)$ is the additive noise. Here we used again the  Born approximation. The calculation of the two-point CINT function 
$\cI(\by,\by')$  from the response matrix $\big(R_{n',n}(t)\big)_{n,n'=0}^N$  is like in \eqref{eq:2CINT}, with the following differences: The matching is done for each source-receiver pair, using the field 
\begin{align*}
\hat{F}_{n',n}(\by,\om) = \hat s(\om) \hat G_o(\by,\bx_n,\om) \hat G_o(\by,\bx_{n'},\om)
\exp\Big( - \frac{x_n^2 + x_{n'}^2}{a^2}\Big),
\end{align*}
and the sensor offset thresholding is done for both the source and receiver pairs. We refer to \cite{borcea2006adaptive,borcea2011enhanced} for the study of the CINT imaging function $\cI^{\CINT}(\by)= \cI(\by,\by)$
in this setup. Using the expressions of the statistical moments of the Green's function, 
it follows after some calculation that the expression of $\EE[\cI(\by,\by')]$  is like that in  \eqref{eq:2.1} and the 
statistical stability is ensured provided that the condition
$\cX \lesssim \cX_d$, $\Omega \lesssim \Omega_d$ is satisfied.

\section{Spectral approach to imaging}
\label{sect:spectral}
In this section we study the linear integral operator $\cL$ with kernel given by the two-point CINT function \eqref{eq:2CINT},
and show that its leading eigenfunction can be used to image the reflectivity 
of a group of point-like reflectors. For simplicity, the imaging is carried out in a range-bin i.e., it is one-dimensional, in the cross-range direction. However, it is possible to extend the  imaging to the range and cross-range plane, as explained briefly at the end of the section. We assume that the threshold parameters are chosen as explained in section \ref{sect:stab},
so we can approximate $\cI(\by,\by')$ by its expectation. The additive noise is ignored in the spectral analysis of $\cL$, although it could be taken into account using the  tools in \cite{kato2013perturbation}. Nevertheless, the numerical simulations in section \ref{sect:num} are for noisy data.  

Before we begin, let us explain how our approach differs from other imaging methods that use
spectral analysis.  The popular MUSIC (Multiple Signal 
Classification) method, which is related to linear sampling \cite{cheney2001linear} and also the factorization method \cite{kirsch2008factorization},  uses the spectral decomposition of the response matrix at a single frequency, collected by an 
active array of sensors to locate small reflectors \cite{lev2000time,ammari2005music}. It can be modified to handle strong additive noise as shown in \cite{borcea2016robust}, but it does not work in random media, where cross correlations have emerged as an important tool for mitigating the wave distortion. In  \cite{leibovich2020generalized,leibovich2021} cross correlations are used to calculate a two-point function like \eqref{eq:2CINT}, but since the imaging is through a homogeneous medium, there is no thresholding. Therefore, the  two-point function  is  an outer product, like  \eqref{eq:HOMMed}, and its samples on the search grid define a rank-one matrix whose leading eigenvector can be used to image. Due to the thresholding needed for statistical stability of imaging in random media, our two-point CINT function is not an outer product and the spectral analysis is much more complicated. It reveals that the leading eigenfunction is useful for improving the resolution of imaging only under a sufficient separation condition between the reflectors. Furthermore,  the resolution of the resulting image is surprising: It is of the order $\sqrt{Hh}$, which is better than the CINT resolution $H$ but worse than the ideal resolution $h$.

\subsection{The integral operator and the imaging function}
\label{sect:spectral1}
\label{eq:operL}
Consider a reflectivity of the form 
\begin{equation}
\rho(\bz) = \sum_{j=1}^M \rho_j \delta(\bz-\bz_j),
\label{eq:S1}
\end{equation}
which models a group of $M$ point-like reflectors at  $\bz_j$, with reflectivity $\rho_j \in \RR$, for $j = 1, \ldots, M$.
Because we image in a fixed range bin $\cD_{\zpar}$ at range coordinate $\zpar$,  we assume  that the reflector locations  are $\bz_j = (z_j,\zpar)$  and the search points in $\cD_{\zpar}$ are $\by = (y,\zpar)$ and $\by' = (y',\zpar)$.

The proposed spectral based imaging function is 
\begin{equation}
\cI^{\SP}(\by = (y,\zpar)):= V_0^\cI(y),
\label{eq:S14}
\end{equation}
where $V_0^\cI(y)$ is the leading eigenfunction of the integral operator ${\cal L}^{\cal I}$ with kernel ${\cal I}(\by= (y,\zpar),\by'= (y',\zpar))$:
$$
\cL^\cI:L^2(\RR) \mapsto L^2(\RR), \qquad \cL^\cI \varphi(y) = \int_{\RR} d y' \, {\cal I}((y,\zpar),(y',\zpar)) \varphi(y'), \qquad \forall \, \varphi \in L^2(\RR).
$$
This kernel is Hermitian and square integrable, 
so  $\cL^\cI$ is self-adjoint and Hilbert-Schmidt  \cite[Chapter 6]{brezis2010functional}.  We also show in 
Appendix \ref{ap:B1} that it is positive semidefinite. 
Our goal in the remainder of the section is to explain when and how the imaging function $\cI^{\SP}$ works.

Substituting \eqref{eq:S1} into the expression of the expectation of the two-point CINT  function obtained from  \eqref{eq:2.1}--\eqref{eq:2.2} and using the statistical stability $\cI(\by,\by') \approx \EE \big[\cI(\by,\by') \big]$, we get  
$$
\cI(\by,\by')  \approx  \mI(y,y') ,
$$
with
\begin{align}
  \mI(y,y') &:= C \sum_{j,j'=1}^M\rho_j \rho_{j'} K_H\Big(\frac{z_j+z_{j'}}{2} - \frac{y+y'}{2} \Big) 
K_h \big( (z_j-z_{j'}) - (y-y')\big),
\label{eq:S2}
\end{align}
where $K_H$ and $K_h$ are defined as in \eqref{eq:kernels}, $H$ and $h$ are given in \eqref{eq:resH}--\eqref{eq:resh}, and $C$ is the redefined constant equal to what we had in 
\eqref{eq:2.2}, divided by $2 \pi \Hpar \hpar$.
The linear integral operator that we wish to study is 
\begin{equation}
\cL:L^2(\RR) \mapsto L^2(\RR), \qquad \cL \varphi(y) = \int_{\RR} d y' \, \mI(y,y') \varphi(y'), \qquad \forall \, \varphi \in L^2(\RR).
\label{eq:S4}
\end{equation}
Again it can be verified that $\cL$ is self-adjoint, Hilbert-Schmidt, and positive semidefinite. The spectral theorem \cite[Chapter 6]{brezis2010functional} then gives that the eigenfunctions $\left(V_n(y)\right)_{n \ge 0}$ of $\cL$ form an orthonormal basis of $L^2(\RR)$  and the  eigenvalues $\left(\Lambda_n\right)_{n \ge 0}$ are non-negative. These eigenvalues are sorted in decreasing order
and $\lim_{n \to \infty} \Lambda_n = 0$.

\subsection{The spectral decomposition}
\label{sect:spectral2}
It is instructive to consider first the case  of  a single reflector, modeled by the reflectivity $\rho(\bz) = \rho_j \delta(\bz-\bz_j)$, for any given $j = 1, \ldots, M$. The operator \eqref{eq:S4} simplifies for such a reflectivity to
\begin{equation}
\cL_j:L^2(\RR) \mapsto L^2(\RR), \qquad \cL_j \varphi(y) = \int_{\RR} d y' \, \mI_j(y,y') \varphi(y'), \qquad \forall \, \varphi \in L^2(\RR),\label{eq:S5}
\end{equation}
with 
\begin{equation}
\mI_j(y,y') := C \rho_j^2 K_H\Big(z_j - \frac{y+y'}{2} \Big) K_h (y'-y).
\label{eq:S6}
\end{equation}
Because the kernel \eqref{eq:S6} is non-negative, $\cL_j$ is a positive operator with non-negative  leading eigenfunction, by the Krein-Rutman theorem \cite{krein1948linear}. For imaging a single reflector we only need this eigenfunction, but the result for the  reflectivity \eqref{eq:S1} uses the full spectral decomposition of $\cL_j$, given in the  next proposition proved in Appendix \ref{ap:B2}.

\vspace{0.03in}
\begin{proposition} 
\label{prop.1} Let $j$ be any index in the set $\{1, \ldots, M\}$.
The eigenvalues of the operator $\cL_j$ are 
\begin{equation}
\Lambda_{j,n} = \frac{C \rho_j^2}{\sqrt{2 \pi} (H + h/2)} \left(\frac{H-h/2}{H+h/2}\right)^n, \qquad n \ge 0,
\label{eq:S7}
\end{equation}
and the eigenfunctions are given by 
\begin{equation}
V_{j,n}(y) = C_{j,n} \exp \Big[-\frac{(y-z_j)^2}{2 Hh} \Big] p_n \Big(\frac{y-z_j}{\sqrt{Hh}}\Big), \qquad n \ge 0,
\label{eq:S8}
\end{equation}
for polynomials $p_n(\xi)$ of degree $n$, 
\begin{equation}
p_n(\xi) = \xi^n + \sum_{l=0}^{n-1} \gamma_{nl} \xi^l,
\label{eq:S9}
\end{equation}
with $p_0(\xi) = 1$ and with coefficients $\gamma_{nl}$  given explicitly in Appendix \ref{ap:B2}. 
Here $C_{j,n}$ is a normalizing constant so that 
$\|V_{j,n}\|_{L^2(\RR)} = 1$.
\end{proposition}

\vspace{0.03in} We conclude from \eqref{eq:S8} that the leading eigenfunction 
\begin{equation}
V_{j,0}(y) = \frac{1}{\sqrt{2\pi Hh}} \exp \Big[-\frac{(y-z_j)^2}{2 Hh} \Big],
\label{eq:S10}
\end{equation}
peaks at the location $z_j$ of the reflector, and it is large for $|y-z_j| = O(\sqrt{Hh})$. Thus, 
this eigenfunction gives an imaging function for a scene with a single reflector, with resolution $\sqrt{Hh}$ that is much better than the CINT resolution $H$, but not as good as the ideal resolution $h$.

Now we can state the result for the reflectivity \eqref{eq:S1}, which models $M$ point-like reflectors.
If these reflectors are not sufficiently spaced out, then the image \eqref{eq:S14} peaks near their locations, but it does not have better resolution than CINT, as shown  with numerical simulations in 
section \ref{sect:num}. The separation condition that allows a better localization with $V_0(y)$ is 
\begin{equation}
\zeta:= \min_{1 \le j,l \le M, j \ne l} \frac{|z_j-z_l |}{3 H} > 1.
\label{eq:S11}
\end{equation} 
The next proposition proved in Appendix \ref{ap:B3} states that if  \eqref{eq:S11} holds, then the interaction between the 
reflectors is weak and the  spectral decomposition of $\cL$ is determined by the decomposition of the operators $\cL_j$, given in Proposition \ref{prop.1}.

\vspace{0.03in}
\begin{proposition}
\label{prop.2}
If the point-like reflectors are well separated, as stated in \eqref{eq:S11}, then the spectral decomposition of the 
operator $\cL$ defined in \eqref{eq:S4} is as follows: The eigenvalues are  
\begin{equation}
\Lambda_n = \sum_{j=1}^M \Lambda_{j,n} + O\left(e^{-9 \zeta^2/2}\right),  \qquad n \ge 0, 
\label{eq:S12}
\end{equation}
with terms given in \eqref{eq:S7} and the eigenfunctions are approximated by the sum 
\begin{equation}
V_n(y) = C_n \sum_{j=1}^M \rho_j V_{j,n}(y) + O\left(e^{-9 \zeta^2/2}\right), \qquad n \ge 0, 
\label{eq:S13}
\end{equation}
with terms given in \eqref{eq:S8}. Here $C_n$ is a normalizing constant so that 
$\|V_n\|_{L^2(\RR)} = 1$.
\end{proposition}

\subsection{Discussion}
\label{sect:spectral4}
It may seem peculiar that according to Propositions \ref{prop.1} and \ref{prop.2}, the imaging function 
\begin{equation}
\cI^{\SP}(y,\zpar) = V_0^\cI(y) \approx V_0(y) \approx C_0\sum_{j=1}^M \rho_j \exp \left[ - \frac{(y-z_j)^2}{2 Hh} \right]
\label{eq:S3}
\end{equation}
has resolution $\sqrt{Hh}$ and yet it assumes the  larger separation condition \eqref{eq:S11}. The 
numerical simulations in section \ref{sect:num} show that \eqref{eq:S11} can be improved a bit to  $\zeta \gtrsim 1/3$. Still, the needed separation is of the order of the resolution of CINT, so what do we gain when imaging with $\cI^{\SP}(y)$? There are two clear advantages:

\vspace{0.03in} \noindent
1. The localization of the reflectors with \eqref{eq:S14} is more precise than with CINT and does not require 
deconvolution as  in \cite{borcea2017imaging}. Therefore, \eqref{eq:S14} is more robust to random fluctuations 
of the two-point CINT function about its mean and also to additive noise.

\vspace{0.03in} \noindent
2. The CINT imaging approach cannot distinguish between positive and negative reflectivities. This can be seen 
from the expression \eqref{eq:S2} evaluated on the diagonal
\begin{align}
\cI^\CINT(\by= (y,\zpar)) &=  \cI(\by,\by) \nonumber \\
&\approx C \sum_{j,j'=1}^M \rho_j \rho_{j'} \cK_H\Big(\frac{z_j+z_{j'}}{2} - y \Big) 
\cK_h (z_j-z_{j'}),
\end{align}
which simplifies under the assumption \eqref{eq:S11} to
\begin{align}
\cI^\CINT(\by ) \approx \frac{C}{2 \pi  h H} \sum_{j=1}^M \rho_j^2  \exp \Big[-\frac{(y-z_j)^2}{2 H^2} \Big].
\label{eq:S16}
\end{align}
Compare this with  \eqref{eq:S3}, which in addition to better localization of the reflectors, it distinguishes the sign of their reflectivity.

We end  with the remark that although we considered imaging at fixed range $\zpar$,
the spectral approach can be extended  to imaging in range and cross range.  This extension is  needed
even for a fixed range bin $\cD_{\zpar}$, because arrival times can be estimated  up to  $O(1/\Omega)$ precision, and thus the points $\by$ in  $\cD_{\zpar}$ are at coordinates  $\by = \big(y,\zpar + O(\Hpar)\big)$.
Indeed, for the reflectivity 
\eqref{eq:S1},  the kernel of the integral operator $\cL$ is 
\begin{align*}
\cI(\by,\by') e^{2 i k_o (\ypar-\ypar')} \stackrel{\eqref{eq:2.1}}{\approx} \cJ(\by,\by') \stackrel{\eqref{eq:2.2}}{=} C \sum_{j,j'=1}^M \rho_{k_o}(\bz_j) \overline{\rho_{k_o}(\bz_{j'})} \\
\times
\cK_{\bH}\Big(\frac{\bz_j + \bz_{j'}}{2} - \frac{\by+\by'}{2} \Big) \cK_{\bh} \big((\bz_j-\bz_{j'})-(\by-\by')\big),
\end{align*}
 with $\rho_{k_o}(\bz_j) = \rho_j \exp(-2 i k_o z_{\parallel,j}),$ for  $j = 1, \ldots, M.$ It is easily verified that this $\cL$ is
 also self-adjoint, positive semidefinite and Hilbert-Schmidt. Moreover, the operator $\cL_j$ associated to a single reflector 
 at $\bz_j$ has a separable kernel in range and cross range
 \[
\cJ_j(\by,\by') = C \rho_j^2 \cK_{\bH}\Big(\bz_j - \frac{\by+\by'}{2} \Big)  \cK_{\bh} (\by-\by'),
\]
so its spectral analysis can be carried out in the two coordinates as we have done in this section. Then, 
under a sufficient separation condition between the reflectors, the spectrum of the operator with kernel 
$\cJ(\by,\by')$ can be written in terms of the spectra of the operators $\cL_j$, similar to Proposition \ref{prop.2}.
 
\section{Optimization approach to imaging}
\label{sect:Wigner}
In this section we describe the second approach for obtaining an image from the two-point CINT function \eqref{eq:2CINT}. It works for a general reflectivity  $\rho(\by)$ of compact support, not just for point reflectors,  as long as  $|\rho(\by)|$ is small enough so that the Born approximation data model \eqref{eq:F9}  is adequate.

Like in the previous section, we assume 
that $\cI(\by,\by')$ is computed using proper thresholding, so we can approximate it in the noiseless case by its expectation
given in  \eqref{eq:2.1}--\eqref{eq:2.2} 
 \begin{align}
\cI(\by,\by')  \approx C e^{-2 i k_o(\ypar-\ypar')}\int_{\cD} d \bz \int_{\cD} d \bz'  \rho_{k_o}(\bz) \overline{\rho_{k_o}(\bz')}\cK_{\bH}\left( \frac{\bz+\bz'}{2}
-\frac{\by+\by'}{2} \right) \nonumber \\
\times  \cK_{\bh}\left( (\bz-\bz')-(\by-\by') \right), \label{eq:W1}
\end{align}
where $\rho_{k_o}(\bz)$ is the modulated reflectivity defined in \eqref{eq:2.4}. 
The noise adds speckle to the image, as explained in section \ref{sect:stab}, and we comment at the end of the section how 
it can be mitigated, at the expense of resolution.

\subsection{The case of noiseless data}
\label{sect:Wig1}
Let us multiply  \eqref{eq:W1} by $\exp[2 ik_o(\ypar-\ypar')]$ and take the Fourier transform with respect to the center point $(\by+\by')/2$ and the offset 
$\by-\by'$. We obtain using 
\begin{equation}
\hat{\rho}_{k_o}(\bK)= \int_{\cD} d \bz \, \rho_{k_o}(\bz) e^{-i \bK \cdot \bz} =  \int_{\cD} d \bz \, \rho(\bz) e^{-i \bK \cdot \bz - 2i k_o \zpar},
\label{eq:W3}
\end{equation}
that 
\begin{align}
\int_{\RR^2} d {\by} \int_{\RR^2} &d {\by'} \, 
e^{-i \tilde{\bK} \cdot \left(\frac{\by+\by'}{2}\right) - i {\bK} \cdot (\by-\by') +2 i k_o(\ypar-\ypar') } \cI(\by,\by') \nonumber \\
&\hspace{-0.5in}\approx C \hat \rho_{k_o} \Big(\bK + \frac{\tilde{\bK} }{2}\Big) \overline{\hat \rho_{k_o} \Big({\bK} -\frac{\tilde{\bK} }{2}\Big)}  \exp \Big( - \frac{\kappa^2 h^2+\kapar^2 \hpar^2+\tilde \kappa^2 H^2+\tilde \kapar^2 \Hpar^2}{2}
\Big), \label{eq:W2}
\end{align}
where $\bK = (\kappa,\kapar)$ and  $\tilde \bK = (\tilde \kappa, \tilde \kapar)$.  
Thus, we can estimate from the two-point CINT function \eqref{eq:2CINT} the following product of the 
Fourier transform of the reflectivity:
\begin{align}
\hat  \rho_{k_o} \Big(\bK + \frac{\tilde{\bK} }{2}\Big)& \overline{\hat \rho_{k_o} \Big({\bK} -\frac{\tilde{\bK} }{2}\Big)} \approx 
P(\bK,\tilde{\bK}) 
\label{eq:W4a}
\end{align}
for arguments
\[
\bK \in \cS_{\bh} := (-3/h,3/h) \times(-3/\hpar,3/\hpar)
\mbox{ 
and }
 \tilde \bK \in \cS_{\bH} :=  (-3/H,3/H) \times (-3/\Hpar,3/\Hpar),
\]
where 
\begin{align}
P(\bK,\tilde{\bK}) 
:= &C^{-1}\exp\Big( \frac{\kappa^2 h^2+\kapar^2 \hpar^2+\tilde \kappa^2 H^2+\tilde \kapar^2 \Hpar^2}{2}
\Big) \nonumber \\
&\times \int_{\RR^2} d {\by} \int_{\RR^2} d \tilde{\by} \, 
e^{-i \tilde{\bK} \cdot \left(\frac{\by+\by'}{2}\right) - i {\bK} \cdot (\by-\by')+2 i k_o(\ypar-\ypar')}  \cI(\by,\by').
\label{eq:W4}
\end{align}

Note that if we evaluate \eqref{eq:W4a} at $\tilde{\bK} = {\bf 0}$ we obtain the modulus squared of \eqref{eq:W3},
\begin{equation}
\left|\hat  \rho_{k_o} (\bK) \right|^2 \approx P(\bK,{\bf 0}).
\label{eq:W5}
\end{equation}
This was used in \cite{borcea2020high} combined with phase retrieval to obtain  an image of  $\rho(\by)$. Since 
$\rho_{k_o}(\by)$ and $\rho_{k_o}(\by-\bz)$ have the same modulus of the Fourier transform, for any fixed $\bz$, the phase retrieval approach is ambiguous with respect to a global shift of the reflectivity scene. Moreover, unless we have strong prior information about the reflectivity, the imaging is not robust to additive noise and to fluctuations of the two-point CINT function about its mean.   
In \cite{borcea2020high} and in the numerical simulations in section \ref{sect:num} the prior is $\rho(\by) \ge 0$.
Other choices such as sparsity of the support of $\rho(\by)$ \cite{shechtman2015phase} or  knowledge of the support  \cite{fienup1987reconstruction} can also been used, but imaging with such priors appears even more sensitive to noise.

Our proposed approach is to   estimate directly the Fourier transform \eqref{eq:W3} from \eqref{eq:W4a}, without phase retrieval. For noiseless data the estimation could be in principle for $\bK \in  \cS_{\bh}$, but as we explain below and show with numerical simulations in section \ref{sect:num}, the method is more robust if we restrict $\bK$ to a smaller rectangle 
$\mathscr{S} \subset  \cS_{\bh}. $

To see that it is possible to find $\hat \rho_{k_o}(\bK)$ from $P(\bK,\tilde{\bK})$, consider 
a cover of $\mathscr{S}$ with  closed rectangular domains ${\itbf r}_j + \tilde{\mathscr{S}}$
centered at nodes $\{{\itbf r}_j = (r_j,r_{\parallel,j}), ~ j = 1, \ldots, J\}$ of a uniform grid with steps 
$\Delta = 1/H$ in cross range and $\Delta_{\parallel} =1/\Hpar$ in range. Here
$\tilde{\mathscr{S}} \subseteq \cS_{\bH}$.  Then, we can estimate recursively $\hat \rho_{k_o}(\bK)$ in each rectangle, starting from the one centered at the origin, as follows:  

\vspace{0.03in}
\textbf{Step 1:} Evaluate  \eqref{eq:W4} at  $\bK = \tilde{\bK}/2$, to get
\begin{equation*}
\hat \rho_{k_o}(\tilde{\bK}) \overline{\hat \rho_{k_o}({\bf 0})} \approx P \left(\frac{\tilde{\bK}}{2},\tilde{\bK}\right), 
\qquad \tilde \bK \in \tilde{\mathscr{S}}.
\end{equation*}
If $\hat \rho_{k_o}({\bf 0}) \ne 0$, which is generically the case, we can determine $\hat \rho_{k_o}(\tilde{\bK})$ for $\tilde \bK \in \tilde{\mathscr{S}}$, up to a multiplicative constant. Note that since $\rho_{k_o}(\by)$ has compact support, its Fourier transform is analytic by the Paley-Wiener theorem, so even if 
 $\hat \rho_{k_o}({\bf 0}) = 0$, we can find ${\boldsymbol{\epsilon}}$ in a small  vicinity of ${ \bf 0}$ where 
 $\hat \rho_{k_o}({\boldsymbol{\epsilon}}) \ne 0$, and therefore we can estimate $\hat \rho_{k_o}(\tilde{\bK}+{\boldsymbol{\epsilon}})$ up to a multiplicative constant.

\vspace{0.03in}
\textbf{Step 2:} For $q = \pm 1, \pm 2, \ldots, $ evaluate  \eqref{eq:W4} at $\bK = \tilde{\bK}/2 + (q\Delta,0)$, to get 
\begin{equation*}
\hat \rho_{k_o}\big(\tilde{\bK}+(q\Delta,0)\big) \overline{\hat \rho_{k_o}\big((q\Delta,0)\big)} \approx P \left(\frac{\tilde{\bK}}{2}
+(q\Delta,0),\tilde{\bK}\right),  \qquad \tilde \bK \in \tilde{\mathscr{S}}.
\end{equation*}
Using the estimated $\hat \rho_{k_o}\big((q\Delta,0)\big)$ at the 
previous steps, determine $\hat \rho_{k_o}\big(\tilde{\bK}+(q\Delta,0)\big)$. Stop when $2(|q|+1)\Delta$ exceeds 
the cross-range side length of  $\mathscr{S}$. 

\vspace{0.03in}
\textbf{Step 3:} After Step 2, we have estimated $\hat \rho_{k_o}(\bK')$ for $\bK' = (\kappa',\kapar')$ in the thin strip 
$|\kapar'| < 3/\Hpar$ in $\mathscr{S}$. Let $q = 0, \pm 1, \pm 2, \ldots, $ and evaluate  \eqref{eq:W4} at 
$\bK = \bK' + \frac{\tilde \bK}{2}+ (0,q \Delta_{\parallel})$ and $\tilde{\bK} = (0,\tilde \kapar)$ to get 
\begin{equation*}
\hat \rho_{k_o}\big(\bK'+(0,\tilde \kapar + q \Delta_{\parallel})\big) \overline{\hat \rho_{k_o}\big(\bK' + (0,q \Delta_\parallel)\big)} \approx P \left( \bK' + \big(0,\frac{\tilde{\kapar}}{2}+q \Delta_{\parallel} \big),
(0,\tilde \kapar)\right).
\end{equation*}
Using the previously estimated $\hat \rho_{k_o}\big( \bK' + (0,q \Delta_\parallel) \big)$, determine $\hat \rho_{k_o}\big(\bK'+(0,\tilde \kapar + q \Delta_{\parallel})\big) $. Stop when $2(|q|+1) \Delta_{\parallel}$ exceeds the range side length of $\mathscr{S}$.

\vspace{0.03in}
This algorithm has the drawback that it involves divisions by complex numbers that may be small.
In our numerical simulations we have found that a more robust estimate  of $\hat \rho_{k_o}(\bK)$ can be obtained with optimization: Since the modulus  $|\hat \rho_{k_o}(\bK)|$ is determined by \eqref{eq:W5}, the optimization involves only
the phase: 
\begin{align}
\min_{\theta(\bK) \in [-\pi,\pi)} \int_{\mathscr{S}} d \bK \int_{\tilde{\mathscr{S}}} d \tilde \bK 
\left| P (\bK,\tilde \bK) - 
\Big| \hat  \rho_{k_o} \Big(\bK + \frac{\tilde{\bK} }{2} \Big) \Big| 
\Big|\hat \rho_{k_o} \Big({\bK} -\frac{\tilde{\bK} }{2}\Big) \Big| \right. \nonumber  \\
\left. \times 
 \exp \Big[ i \theta \Big(\bK + \frac{\tilde{\bK} }{2}\Big) - i \theta \Big(\bK - \frac{\tilde{\bK} }{2}\Big)\Big] 
\right|^2.\label{eq:W6Opt}
\end{align}
The minimization \eqref{eq:W6Opt} may be expensive, but in our experience it is feasible to use at least for 
one range bin at a time, as shown in section \ref{sect:num}. Although we have not done so, the problem can 
be regularized by  adding to the objective function some penalty term like the $L^1(\cD)$ or $TV$ norm of $\rho(\by)$. 

If we denote by $\theta^{\rm est}(\bK)$ the estimated phase, then we have 
\begin{equation}
\hat\rho_{k_o}^{\rm est}(\bK):= |\hat \rho_{k_o}(\bK)| \exp \big[ i \theta^{\rm est}(\bK)\big] = 
\sqrt{P(\bK,{\bf 0})}  \exp \big[ i \theta^{\rm est}(\bK)\big],
\end{equation}
and we can image the reflectivity using the inverse Fourier transform
\begin{equation}
\cI^{\OP}(\by):= e^{2 i k_o \ypar} \int_{\RR^2} \frac{d \bK}{(2 \pi)^2} \, e^{i \bK \cdot \by} \chi_{_{\mathscr{S}}}(\bK) \hat \rho_{k_o}^{\rm est}(\bK).
\label{eq:W6}
\end{equation} 
Here $\chi_{_{\mathscr{S}}}(\bK)$ is a smooth tapering function supported in the estimation domain $\mathscr{S}$. 
The Fourier transforms are carried out with the FFT algorithm, and so the integrals in \eqref{eq:W6Opt} and \eqref{eq:W6} 
are approximated by sums.
\subsection{Noisy data and filtering}
\label{sect:Wig2}
We see from \eqref{eq:W6} that if $\hat \rho_{k_o}(\bK)$ can be estimated robustly in a domain $\mathscr{S}$ as large as 
$ \cS_{\bh}$, then the resolution of the image \eqref{eq:W6} is the ideal one, described by the components of $\bh$. This ideal resolution cannot be achieved in practice due to noise effects, which manifest 
as speckle in the two-point CINT function, with typical size of order $h$ in cross range 
and $\hpar$ in range (recall section \ref{sect:stab}).  The variations of $\cI(\by,\by')$ about its mean also 
affect the result.  The imaging is stabilized by using  a domain $\mathscr{S} = [-1/h^{\rm est},1/h^{\rm est}\big] \times [-1/\hpar^{\rm est},1/\hpar^{\rm est}]$, with 
\[
H \gg h^{\rm est} \gg h, \qquad \Hpar \gg \hpar^{\rm est} \gg \hpar.
\]
The resolution of the image is then given by $\bh^{\rm est} = (h^{\rm est},\hpar^{\rm est})$, which is much better than that of CINT, but not as good as the ideal one given by $\bh = (h,\hpar)$.

\section{Numerical results}
\label{sect:num}
In this section we use numerical simulations to assess the performance of the two imaging approaches proposed in sections \ref{sect:spectral} and \ref{sect:Wigner} and to compare with conventional SAR imaging, CINT imaging and 
imaging using the phase retrieval approach introduced in \cite{borcea2020high}.

\subsection{Setup}
We consider a reflectivity of the form (\ref{eq:S1}), with three or five reflectors in a single range bin, so the imaging is 
one-dimensional, along  the cross-range direction. The data are generated with the random travel model \eqref{eq:F9},
using a random phase screen with correlation length $\ell = a/2$. The reference length scale is the central wavelength 
$\la_o$, the range bin is at distance $L = 2 \cdot 10^4 \la_o$ and the aperture is $a = L/(2 \pi)$. The random phase of the Green's function has mean zero and  variance (see 
\eqref{eq:A6})
\begin{equation}
\EE \left[ \big(\om_o \tau_\mu(\bz,\bx_n) \big)^2\right] =: \sigma_{\tau}^2.
\label{eq:N1}
\end{equation} 
By varying $\sigma_{\tau}$ we increase or decrease the wavefront distortion. The decoherence length 
is, according to \eqref{eq:A12}, 
$\cX_d = {\sqrt{3} \ell}/(2 \sigma_\tau).$
Because we image only in cross range, 
we use time-harmonic data, so the cross correlations are only over the sensor locations, offset by at most $\cX = \min\{a, \cX_d/3\}.$ 
The additive noise is modeled as explained in section \ref{sect:Form.2}, with standard deviation $\sigma_{_W}$. We show results 
with $\sigma_{_W} = 0$ and with $\sigma_{_W}$ such that the noise term in \eqref{eq:F9a} is $10\%$ of the maximum value of the recorded signal.

The aperture is centered at the origin, and consists of $N  = 400$ sensor locations. The imaging domain is the interval
$\cD = (0,245\la_o)$ and it is sampled on a fine uniform grid with spacing $0.03\la_o$. 
The SAR imaging function is computed as in \eqref{eq:ISAR} and the two-point CINT function and the CINT image are computed as in \eqref{eq:2CINT} and  \eqref{eq:CINT}.  The leading eigenvector which defines the  image \eqref{eq:S14} 
is calculated with the power method. The optimization \eqref{eq:W6Opt} is carried out with the MATLAB optimization routine
``fminunc". This ignores the constraint $\theta(\kappa) \in [-\pi,\pi)$ so it gives an estimated phase up to an integer factor of $2 \pi$.  
The phase retrieval image is obtained from the modulus $|\hat \rho(\kappa)|$ given by \eqref{eq:W5} evaluated at $\bK = (\kappa,0)$,  using the standard algorithm \cite{fienup1978reconstruction,shechtman2015phase} with positivity constraint: At each iteration of this algorithm, we  Fourier transform the current estimate of $\rho$ and update the modulus to match the given  $|\hat \rho(\kappa)|$ at $|\kappa| < 3/h$. The higher $|\kappa|$ components are set to $0$. The result is then inverse Fourier transformed and the phase is adjusted to match the positivity constraint $\rho(y) \ge 0$. 

\begin{figure}[t]
\centerline{
\raisebox{0.1in}{\includegraphics[width = 0.33\textwidth]{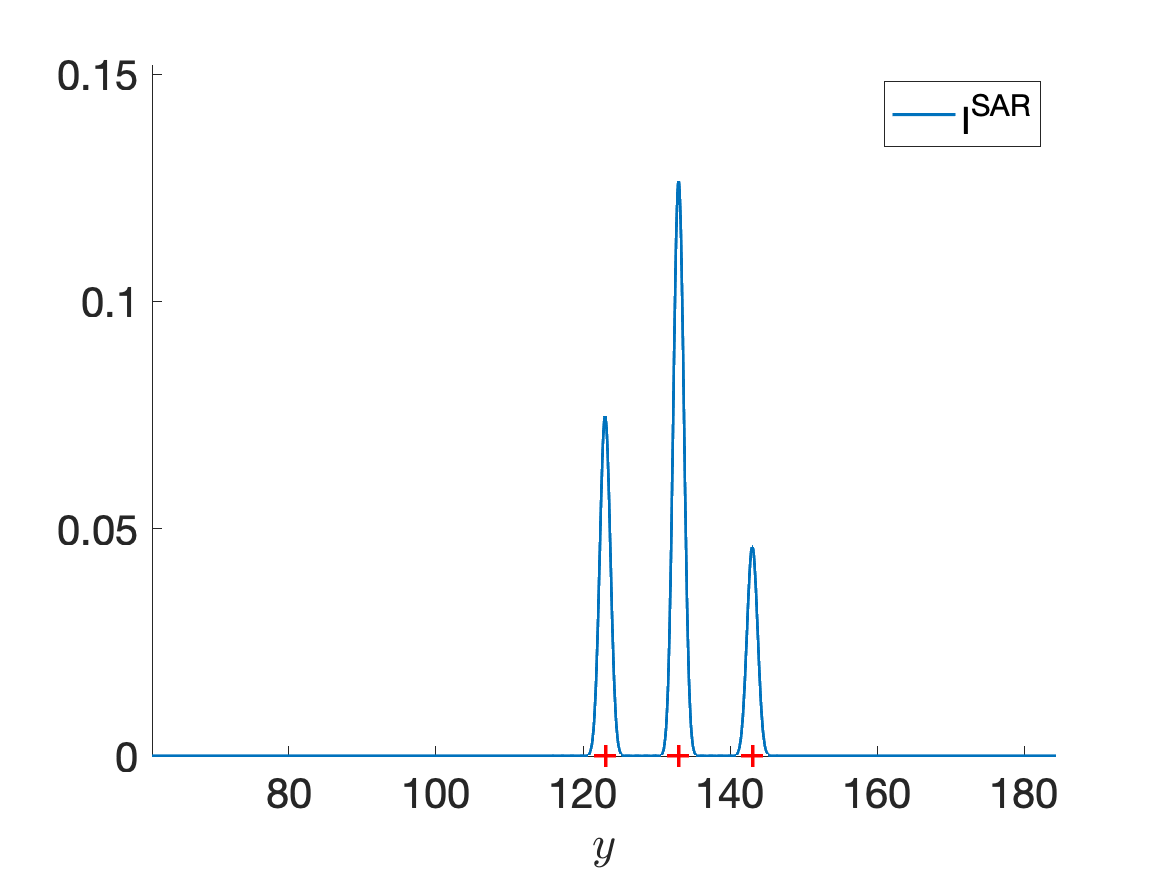}}\includegraphics[width = 0.4\textwidth]{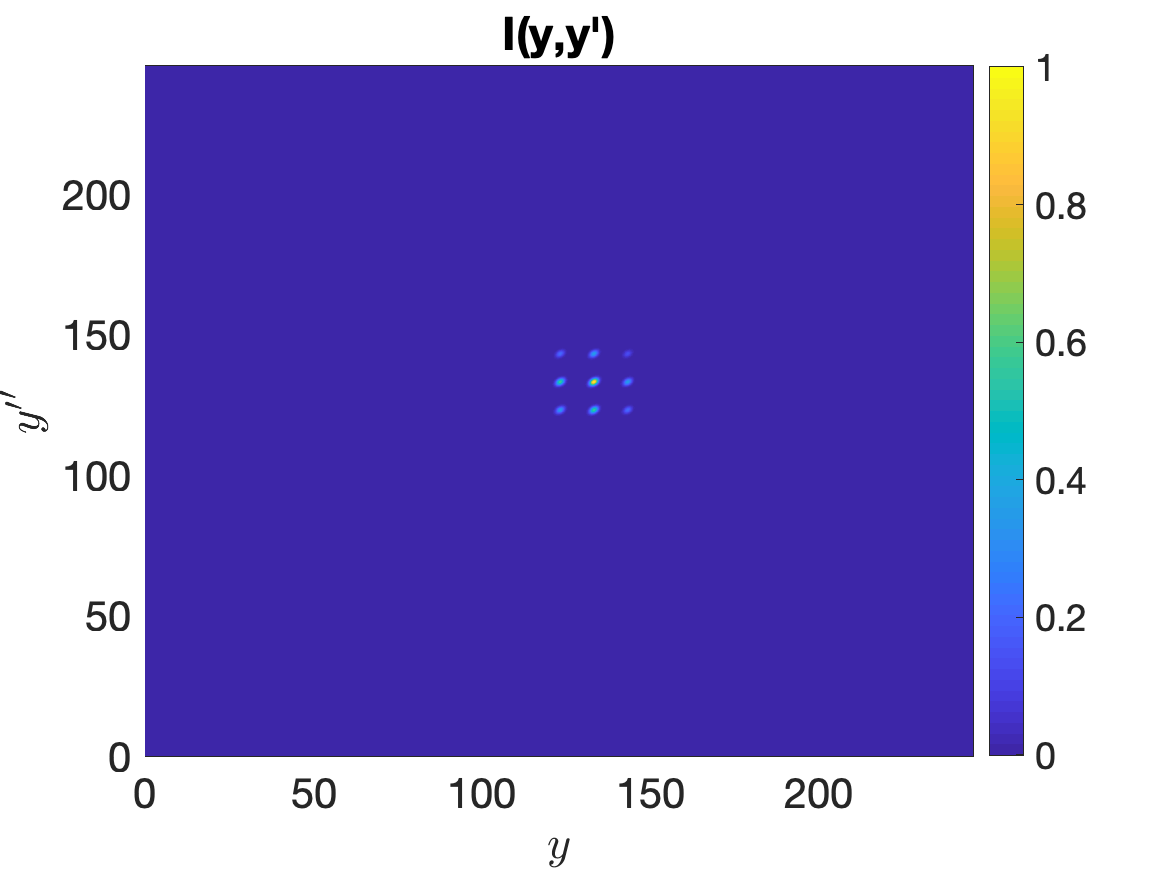}
}
\centerline{
\includegraphics[width = 0.33\textwidth]{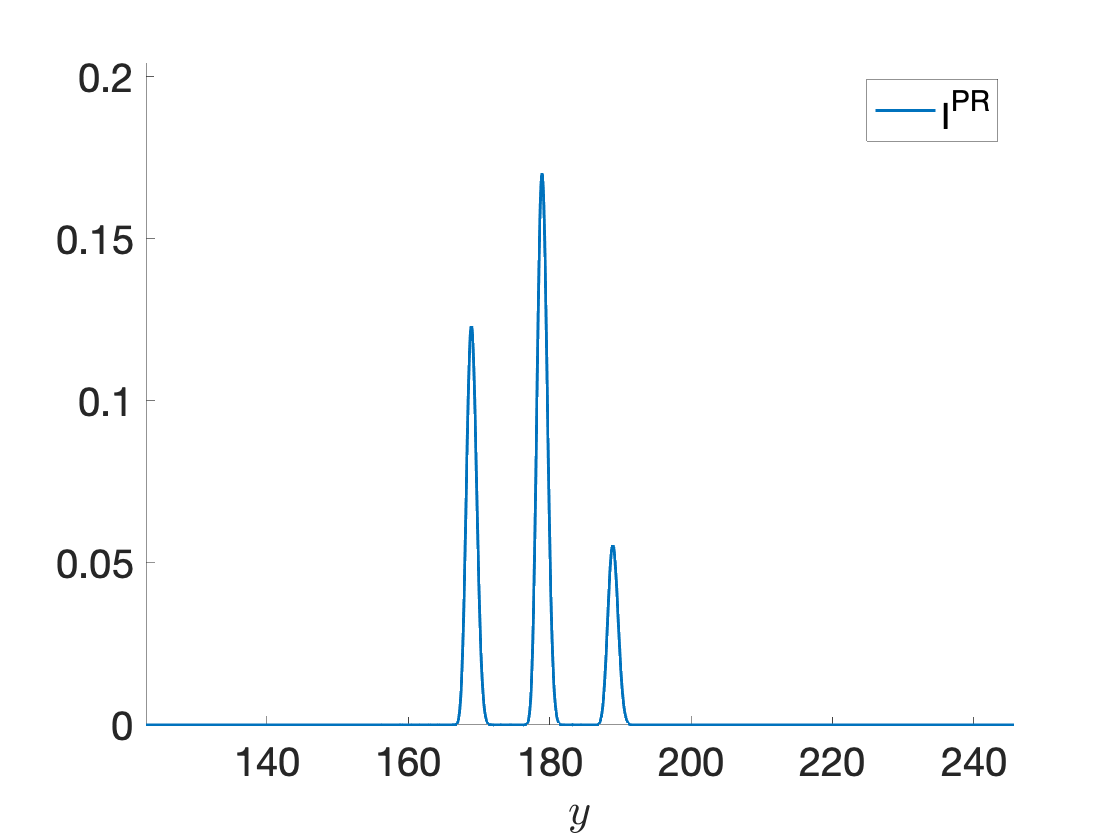}
\includegraphics[width = 0.33\textwidth]{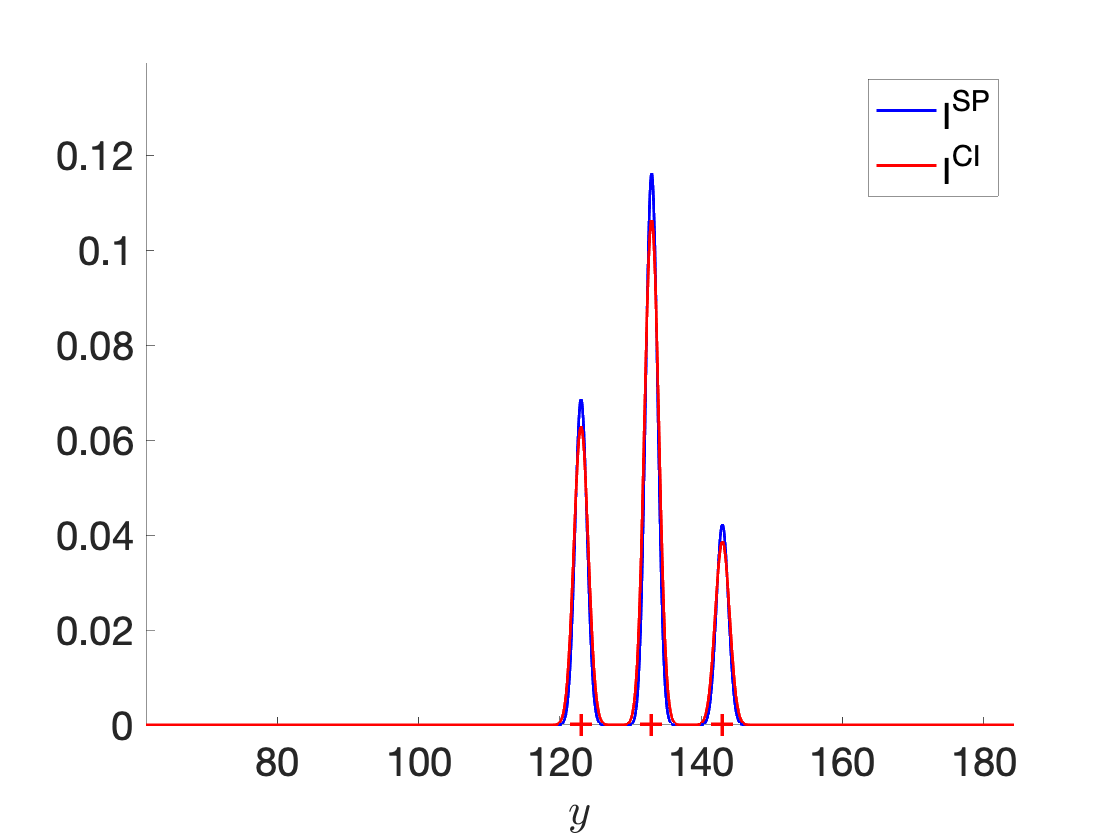}
\includegraphics[width = 0.33\textwidth]{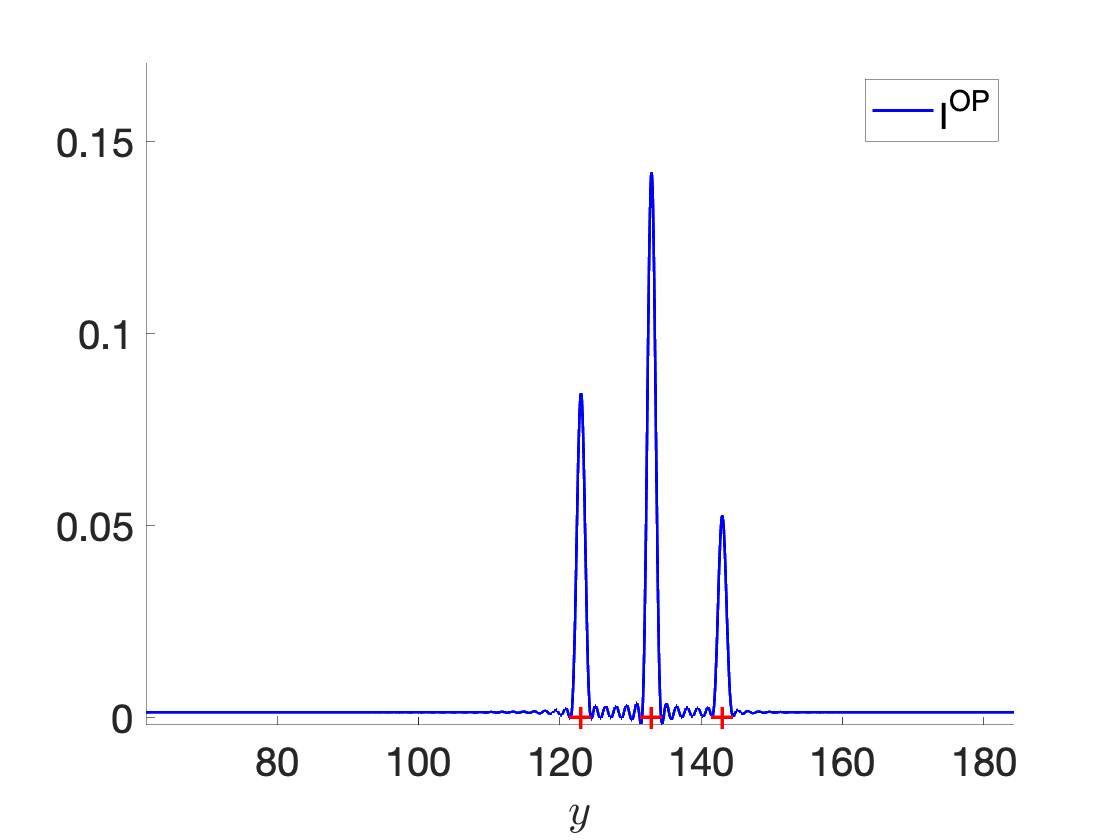}
}
\caption{Results for the reflectivity \eqref{eq:Refl1} obtained with ideal data: noiseless and in the homogeneous medium.  The SAR image $\displaystyle \cI^{\SAR}(y)$ is in the top left plot. The two-point CINT function  $\cI(y,y')$ is in the 
top right plot. The bottom line shows: The image $\cI^{\PR}(y)$ obtained via phase retrieval (left), the square root of the CINT image $\cI^{\CI} (y):= \sqrt{\cI^{\CINT}(y)}$ (red line) and the spectral based image  $\cI^{\SP}(y)$ (blue line) are in the middle plot and the optimization based image
$\cI^{\OP}(y)$ is in the right plot. The axes are in  $\la_o$ units. The locations of the reflectors are indicated with red crosses on the abscissa.}
\label{fig:1}
\end{figure}

\begin{figure}[h]
\centerline{
\raisebox{0.1in}{\includegraphics[width = 0.33\textwidth]{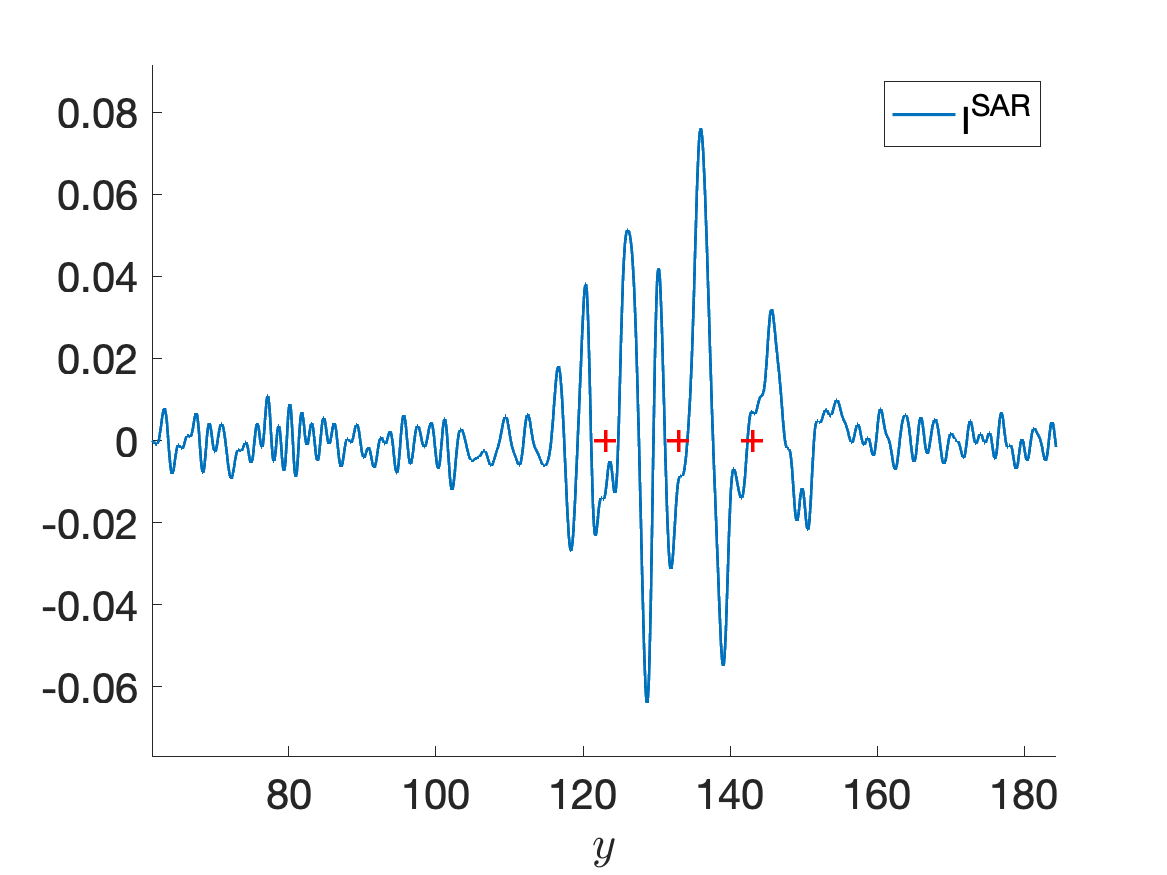}}\includegraphics[width = 0.4\textwidth]{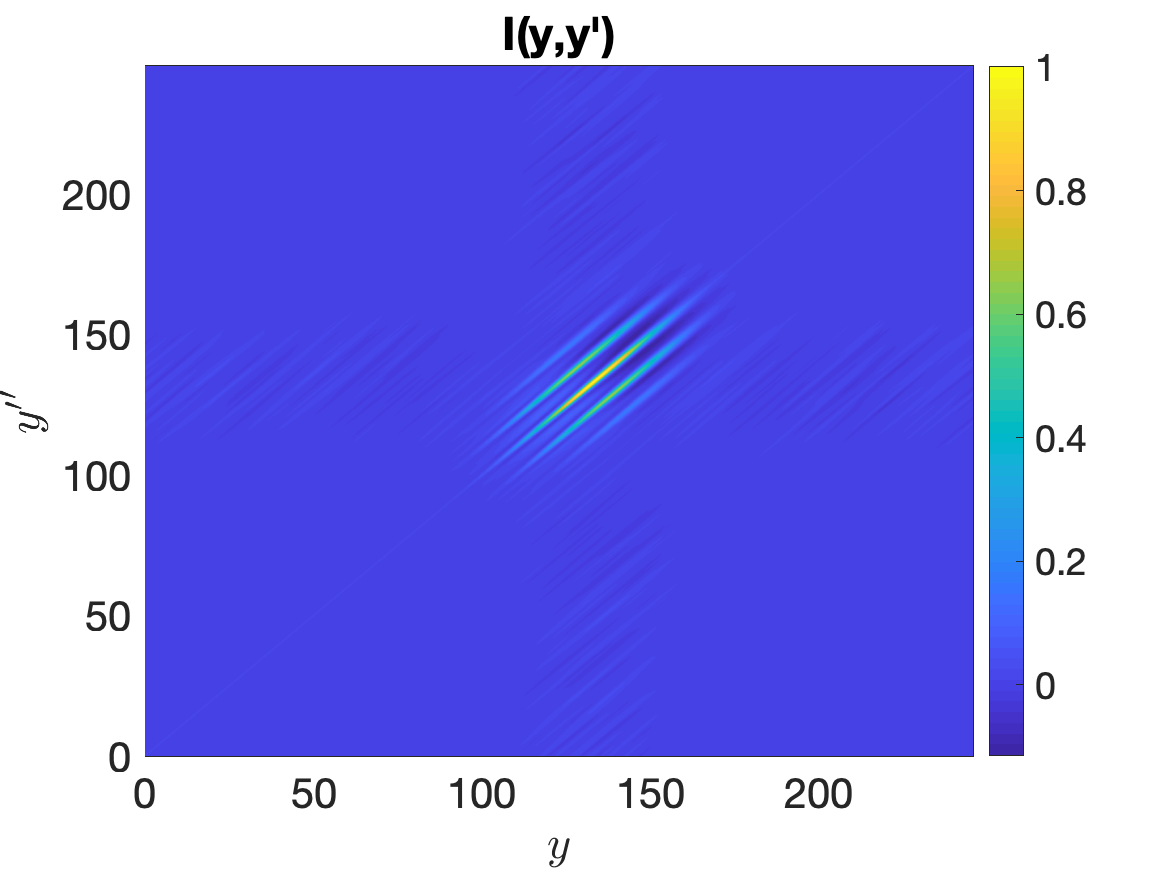}
}
\centerline{
\includegraphics[width = 0.33\textwidth]{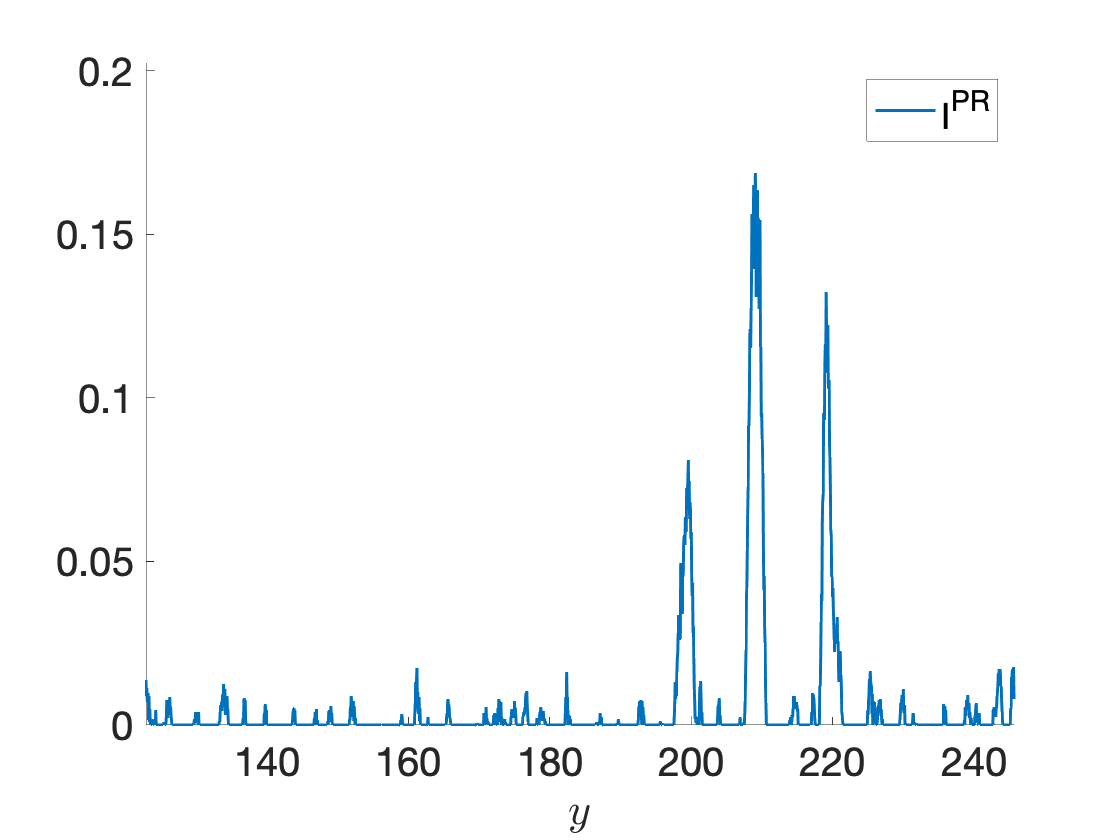}
\includegraphics[width = 0.33\textwidth]{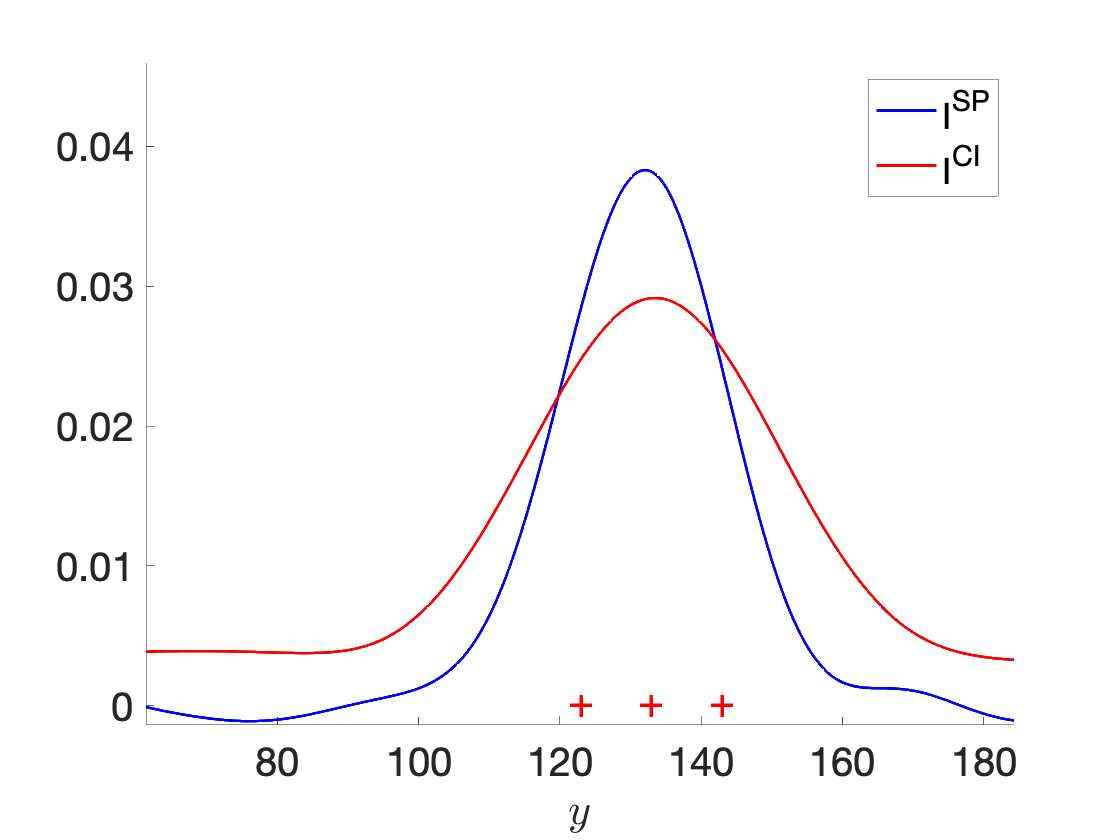}
\includegraphics[width = 0.33\textwidth]{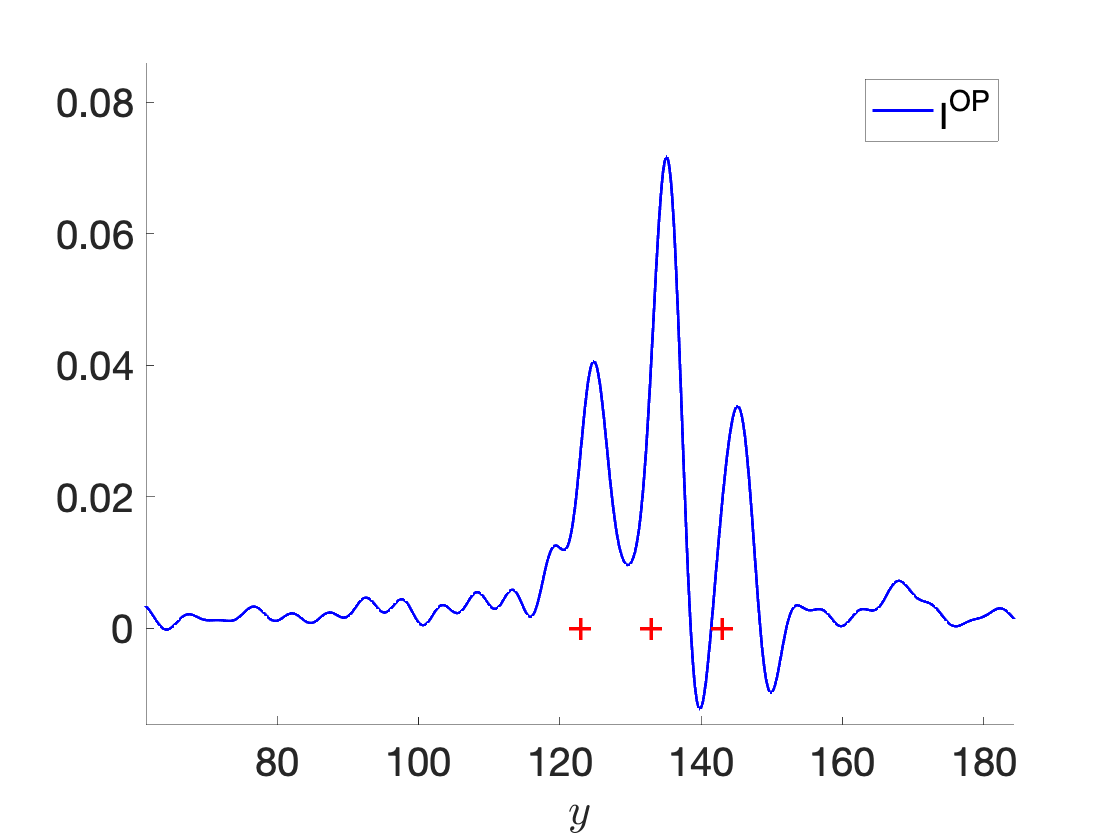}
}
\caption{Results for the reflectivity \eqref{eq:Refl1} in the random medium with $\sigma_\tau = 3.1$ and with $10\%$ additive noise.  
}
\label{fig:2}
\end{figure}

\begin{figure}[h]
\centerline{
\raisebox{0.1in}{\includegraphics[width = 0.33\textwidth]{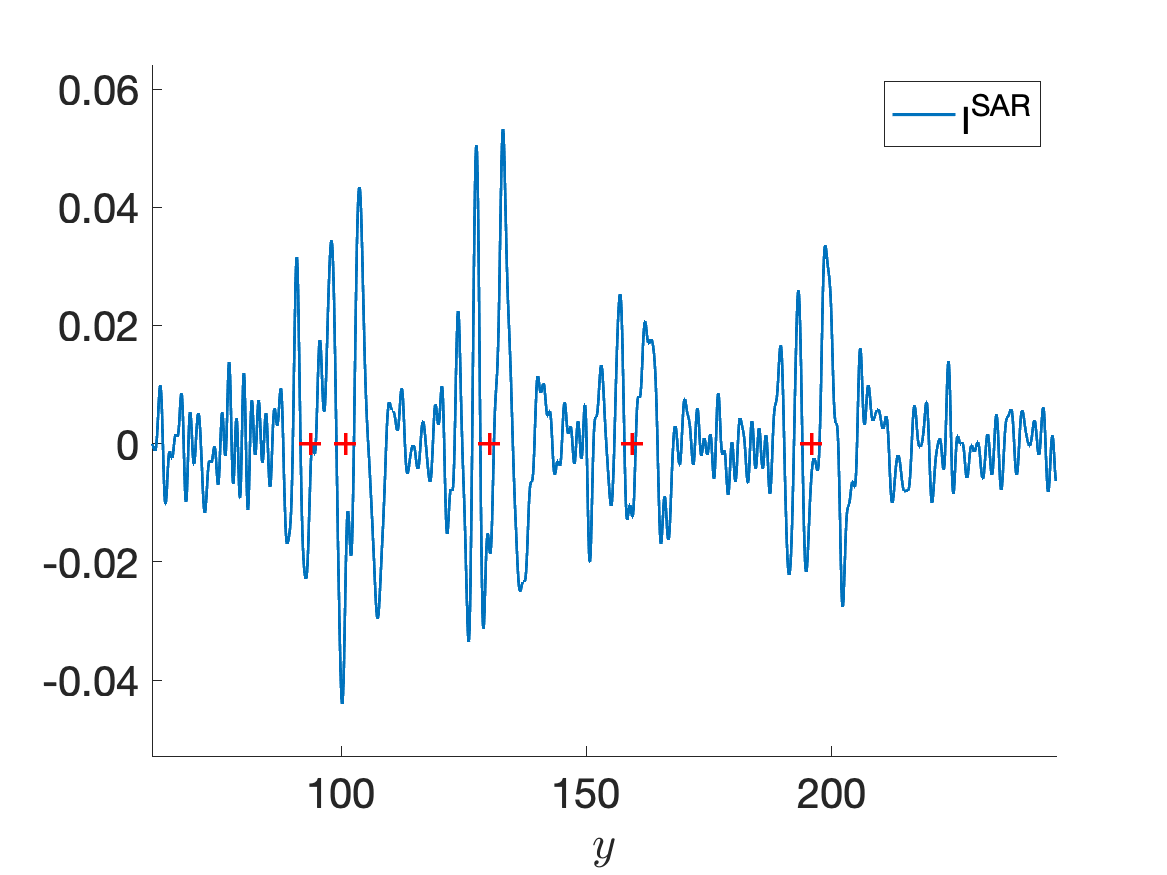}}\includegraphics[width = 0.4\textwidth]{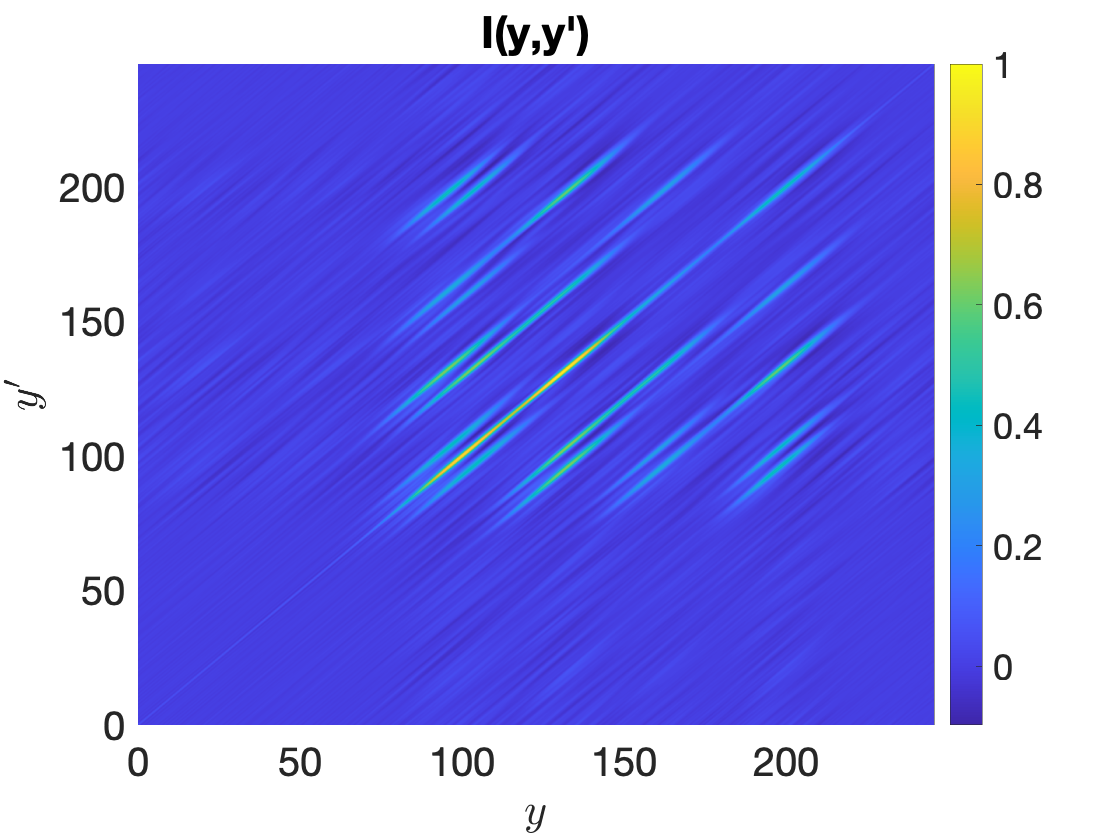}
}
\centerline{
\includegraphics[width = 0.33\textwidth]{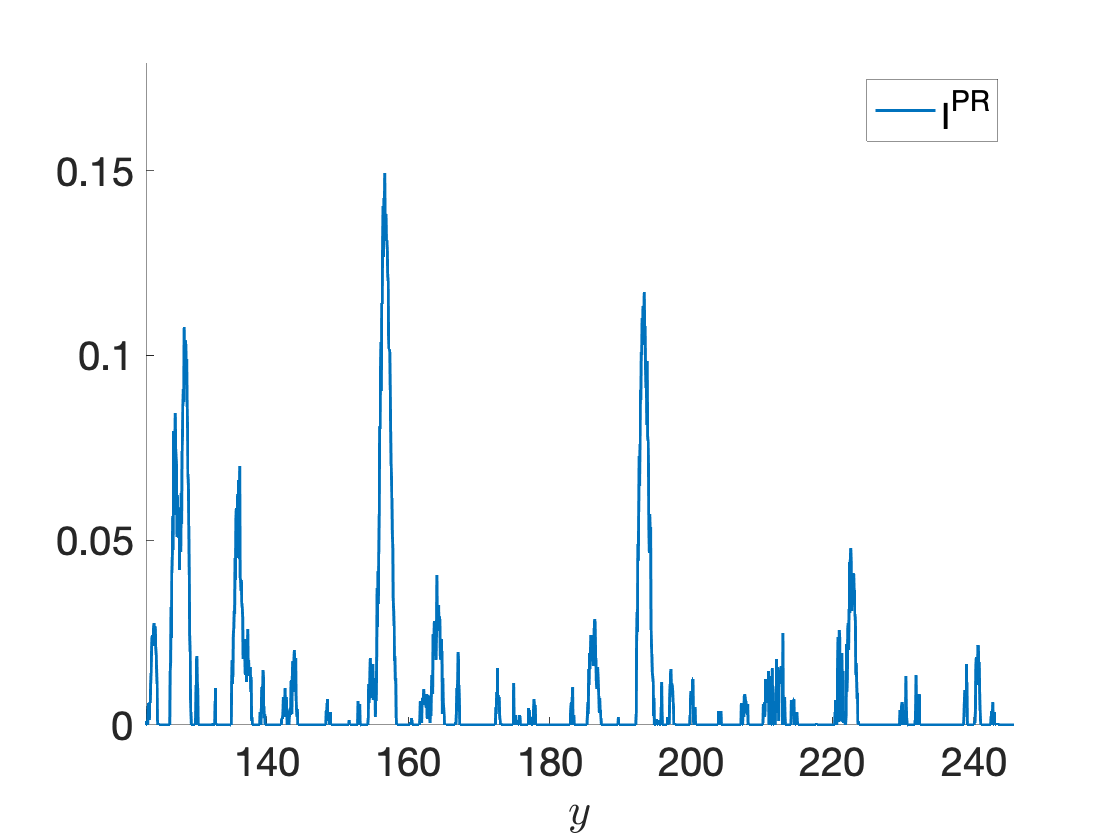}
\includegraphics[width = 0.33\textwidth]{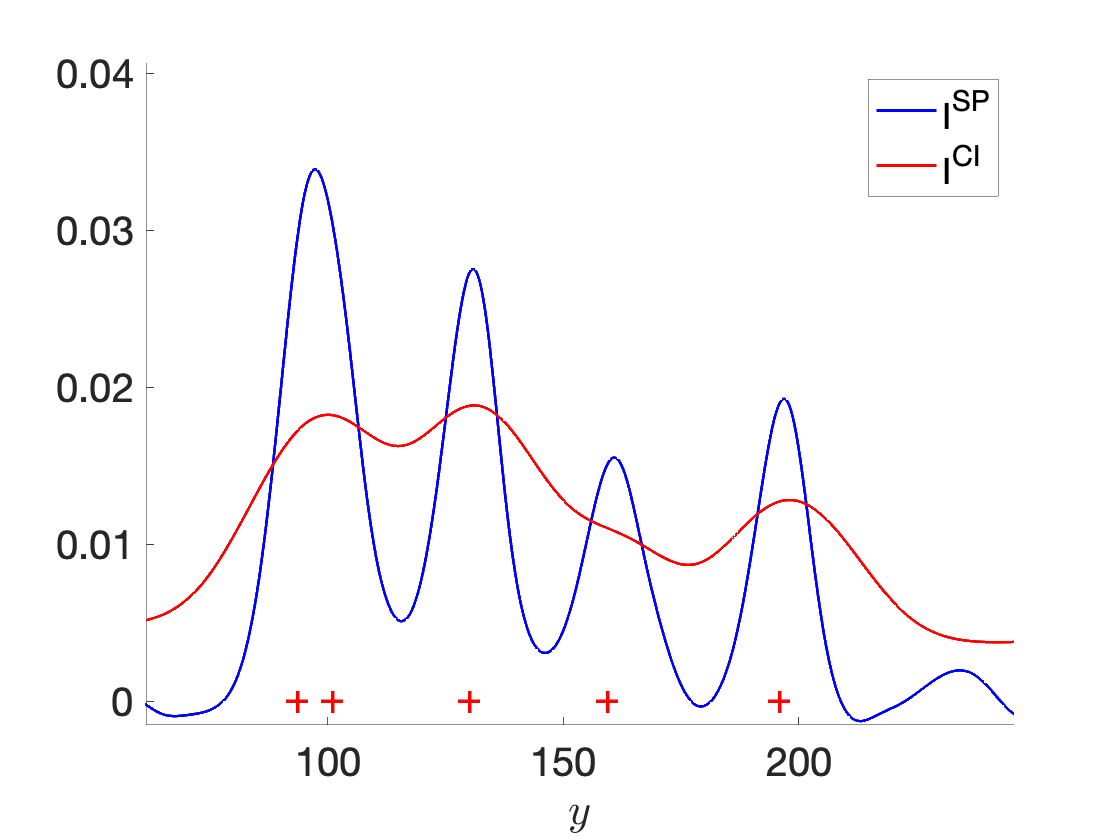}
\includegraphics[width = 0.33\textwidth]{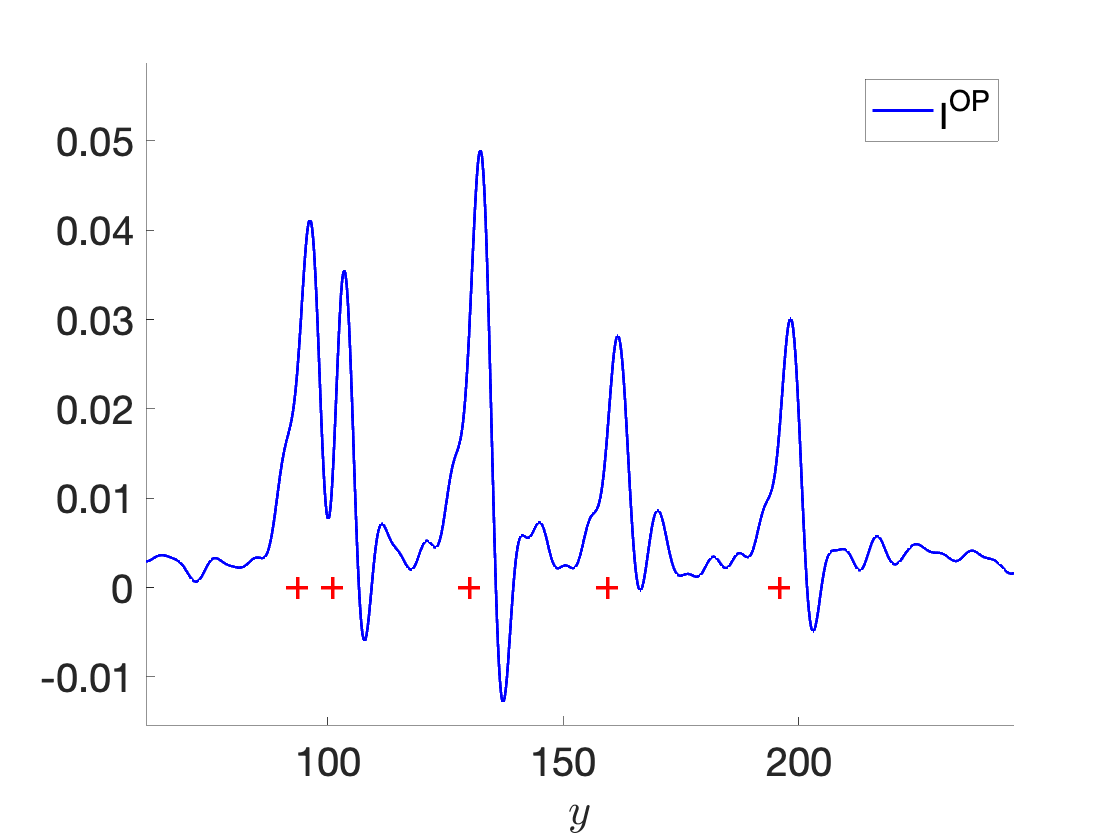}
}
\caption{Results obtained for the reflectivity \eqref{eq:Refl2} in the random medium with $\sigma_\tau = 3.1$ and with $10\%$ additive noise.  
}
\label{fig:3}
\end{figure}

\begin{figure}[h]
\centerline{
\raisebox{0.1in}{\includegraphics[width = 0.33\textwidth]{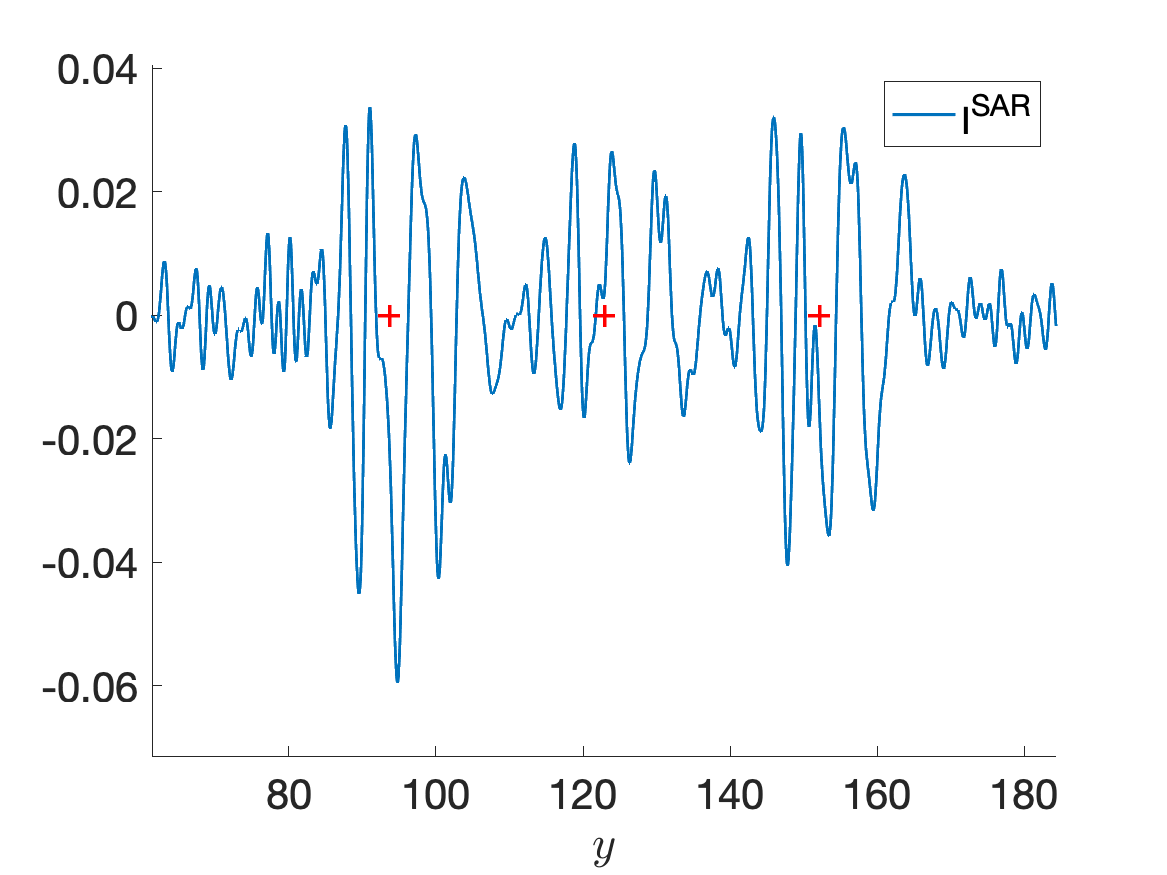}}\includegraphics[width = 0.4\textwidth]{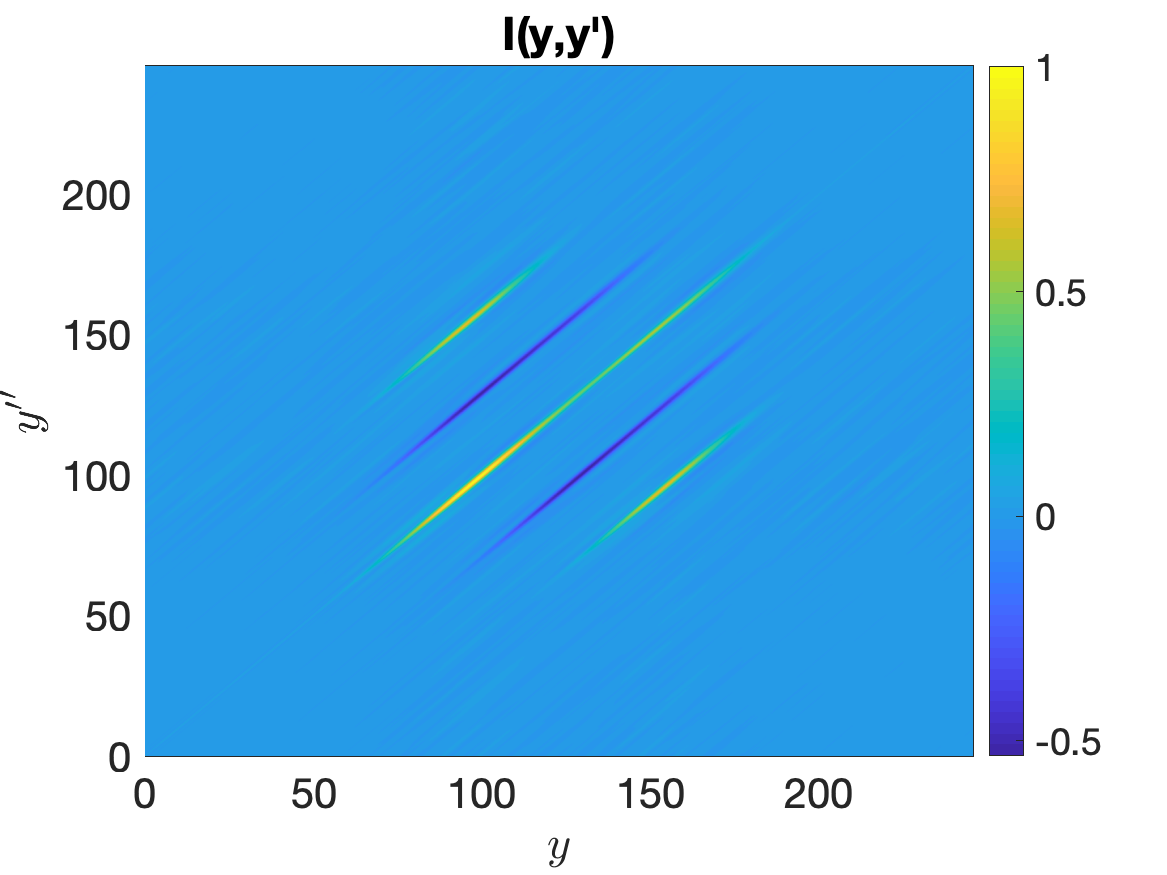}
}
\centerline{
\includegraphics[width = 0.33\textwidth]{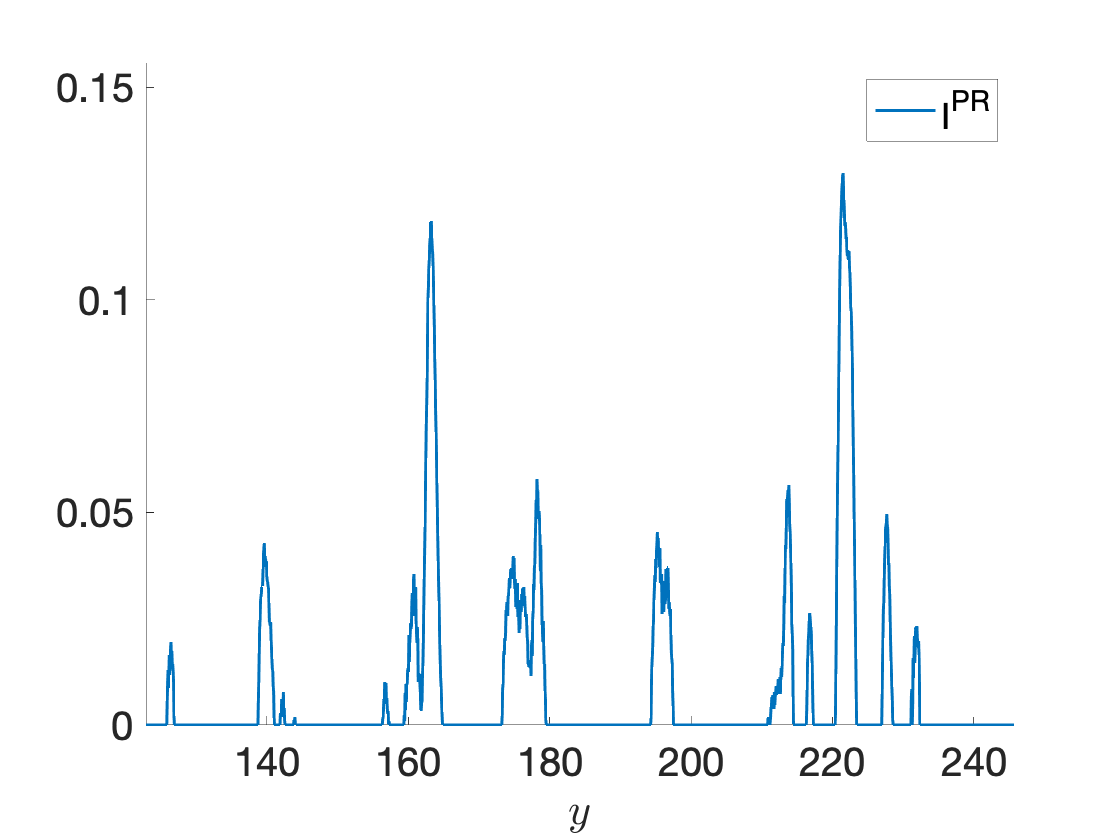}
\includegraphics[width = 0.33\textwidth]{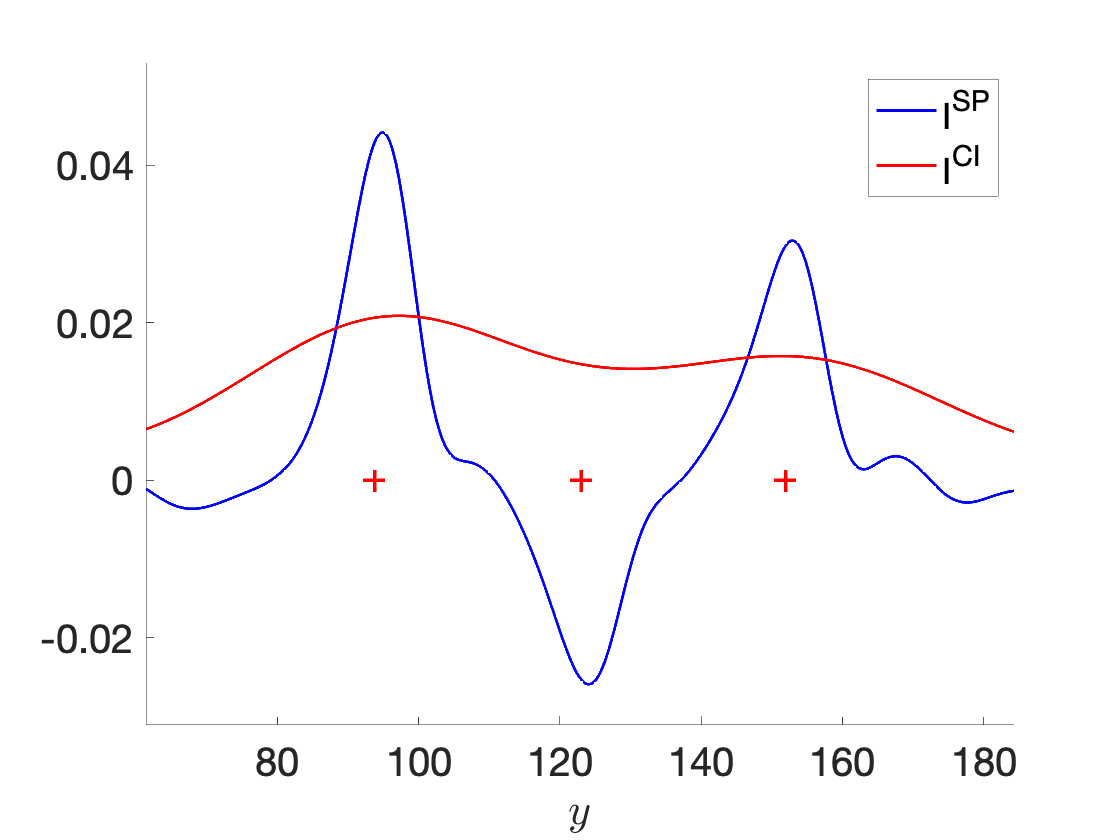}
\includegraphics[width = 0.33\textwidth]{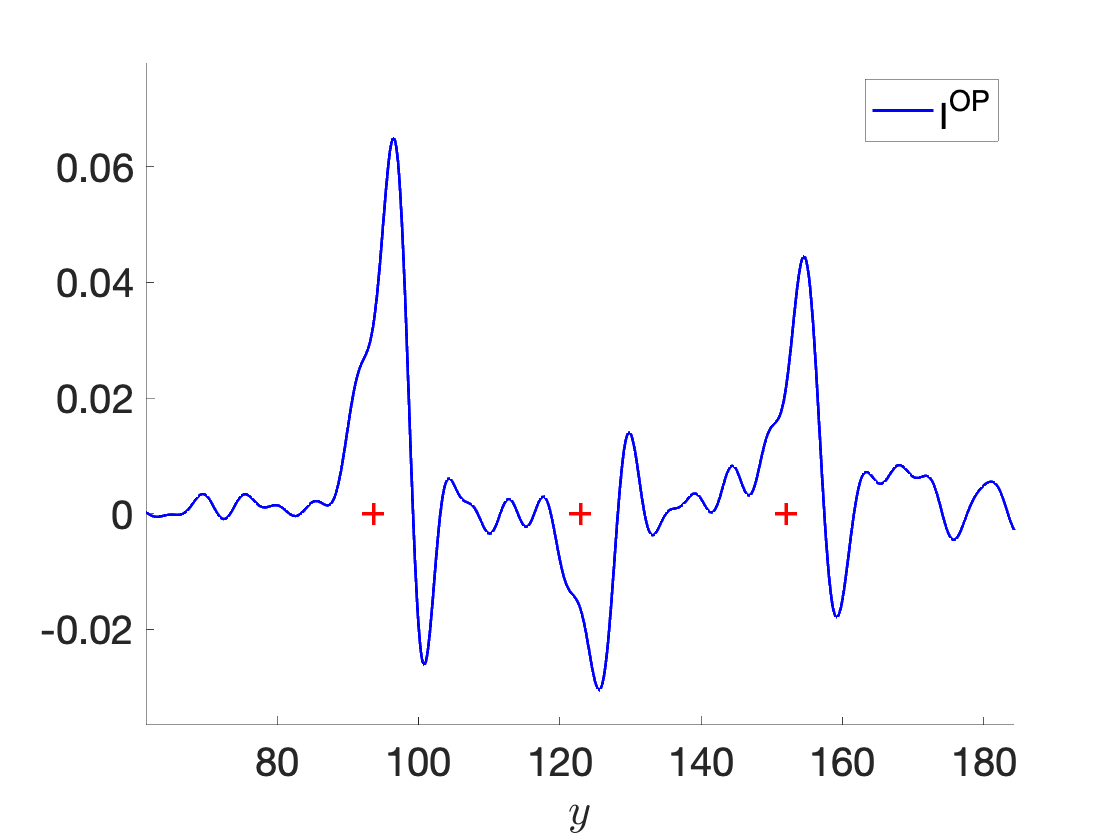}
}
\caption{Results obtained for the reflectivity \eqref{eq:Refl3} in the random medium with $\sigma_\tau = 4$ and with $10\%$ additive noise. 
}
\label{fig:4}
\end{figure}

\subsection{Results}
We begin in Fig. \ref{fig:1} with results obtained in the ideal case where the medium is homogeneous
and there is no noise. There are three reflectors in this simulation, modeled by 
\begin{equation}
\rho(y) = 2.2 \delta(y-133\la_o) + 1.3 \delta(y-123\la_o) + 0.8 \delta(y-143\la_o),
\label{eq:Refl1}
\end{equation}
and we see from the top left plot that the conventional SAR image $\cI^\SAR(y)$ localizes them very well. The two-point CINT function $\cI(y,y')$ is displayed in the top right plot. It has nine distinct peaks, corresponding to  the 
offsets between the coordinates of the reflectors. The plots in the bottom line show the image $\cI^{\PR}(y)$  obtained via phase 
retrieval (left plot), the images $\cI^{\SP}(y)$ and $\cI^{\CI}(y) := \sqrt{\cI^\CINT(y)}$ (middle plot) and $\cI^{\OP}(y)$ 
(right plot). Because the CINT image is squared in the reflectivity, we take its square root to compare it with 
the other functions. We conclude from the plots that all methods give equally good estimates of the reflectivity \eqref{eq:Refl1}, although the phase retrieval image is ambiguous to an overall shift, as expected. 

In Fig. \ref{fig:2} we show the results for the same reflectivity \eqref{eq:Refl1}, except that the data are contaminated with $10\%$ additive noise and they are distorted by a random medium, as quantified by the  standard deviation $\sigma_\tau = 3.1$ (recall \eqref{eq:N1}). We use  a single realization of the medium, so there is no averaging over realizations.
The images are displayed in the same format as in Fig. \ref{fig:1}. Note the significant deterioration of the SAR image,
which exhibits many spurious large peaks. Repeated simulations show that these change unpredictably with the realization of the random medium. The two-point CINT function is insensitive with respect to the realization of the medium,
but as shown by \eqref{eq:S2} it is affected by the blur along the diagonal. The resolution parameters are $H = 11.36\la_o$ 
and $h = \la_o$. The separation between the reflectors is $\zeta = 0.88H$, so the CINT image $\cI^{\CI}(y)$ and the spectral based 
image $\cI^{\SP}(y)$ cannot resolve the reflectors. The phase retrieval  image $\cI^{\PR}(y)$ resolves the reflectors, up to an ambiguous overall shift. The image $\cI^{\OP}(y)$ locates correctly the reflectors, so it has better resolution 
than $\cI^{\SP}(y)$, but not as good as in the homogeneous medium. This is due to the filtering explained in section \ref{sect:Wig2}. This filtering also leads to some Gibbs-like oscillations, which can be mitigated in principle by adding some 
regularization to the optimization that minimizes say the $L^1(\cD)$ norm of $\rho^{\rm est}$. 

In Fig. \ref{fig:3} we show the results for imaging through the same random medium as in Fig. \ref{fig:2} 
and with $10\%$ additive noise, except that the scene has five reflectors
\begin{equation}
\rho(y) =  2 \delta(y-93.7 \la_o) + 2 \delta(y-101\la_o) + 3 \delta(y-130 \la_o) + 1.5 \delta(y-159\la_o) +
2 \delta(y-196 \la_o).
\label{eq:Refl2}
\end{equation}
The first two reflectors are within $0.88H$ distance but the other reflectors are better separated at offsets $2.56H$, 
$2.56H$ and $3.26H$, respectively. As predicted by the analysis in section \ref{sect:spectral}, the image 
$\cI^{\SP}(y)$ locates the well separated reflectors with better resolution than CINT, but it does not resolve the nearby reflectors.
The optimization approach resolves all the reflectors (bottom right plot), where again the Gibbs-like oscillations are due to filtering. The phase retrieval approach gives a much noisier image than in Fig. \ref{fig:2} and the SAR image is not informative.  

Finally, we show in Fig. \ref{fig:4} the results for imaging through a stronger random medium, with $\sigma_\tau = 4$,
and with $10\%$ additive noise.  The resolution scales are $H = 14.615\la_o$ and $h = \la_o$. The scene consists of three reflectors, one with negative reflectivity, 
\begin{equation}
\rho(y) = 2 \delta(y-93.7\la_o) - \delta(y-123\la_o) + 1.5 \delta(y-152\la_o).
\label{eq:Refl3}
\end{equation}
The separation between the reflectors is $2 H$ and $1.98H$, which is slightly smaller than what we assumed in section \ref{sect:spectral}. Still, the spectral based image $\cI^{\SP}(y)$ localizes correctly the reflectors, with 
better resolution than CINT and it also gives correctly the signs and relative magnitudes of the reflectivities. The 
optimization-based imaging method also works well and gives better resolution, aside from the Gibbs-like oscillations. 
The SAR image is useless and the phase retrieval based approach gives also a wrong result because our implementation
assumes a positive reflectivity, which is not the case in this simulation.

\section{Summary}
\label{sect:sum}
In this paper we have shown that the two-point CINT function calculated from the empirical cross correlations of the measured wave field contains the relevant information to obtain statistically stable and high-resolution synthetic aperture radar (SAR) images of the reflectivity of a remote scene.  We  have introduced two new methods to build such images starting from first principles and taking into account additive measurement noise and clutter induced effects i.e., wave distortion due to random scattering. With analysis and numerical simulations, we have characterized the pros and cons of five imaging approaches. The first is the conventional SAR imaging method. The other 
four methods, which include the novel two, are based on the two-point CINT function:

\vspace{0.05in}
\begin{itemize}
\itemsep 0.05in
\item
The conventional SAR imaging function $\cI^{\SAR}(\by)$ is robust to measurement noise but not to clutter induced effects. If these effects are negligible, it is computationally cheap, it has high resolution, it works no matter the 
sign of the reflectivity function $\rho(\by)$, and it can recover the sign of $\rho(\by)$. If clutter induced effects are significant, $\cI^{\SAR}(\by)$ is useless because of strong and unpredictable artifacts.

\item
The standard CINT imaging function 
$\cI^{\CI}(\by) = \sqrt{\cI^{\CINT}(\by)}$ is robust to both measurement and clutter induced effects, but has poor resolution.
It  has moderate computational cost, it  works  with a reflectivity function  that may change sign,  but it cannot recover the sign.

\item The phase-retrieval  imaging function $\cI^{\PR}(\by)$ is robust to clutter induced effects. 
If the reflectivity function has constant sign, it is also robust to measurement noise and it gives a high resolution image. However, the phase retrieval step makes the imaging computationally expensive, and the resulting image 
has an uncertain  global shift and symmetry with respect to the origin. If the reflectivity changes sign, the imaging is 
sensitive to noise.

\item
The spectral based imaging function 
$\cI^{\SP}(\by)$ has moderate resolution,  but it is very robust to measurement noise and to clutter induced effects.
It is computationally moderately expensive,  thanks to the power method, it works with a reflectivity function that may change sign and it can recover the sign of $\rho(\by)$. However, it requires a sparse enough imaging scene, where the 
reflectors are sufficiently well separated.

\item The optimization based imaging function 
$\cI^{\OP}(\by)$ has moderate to high resolution and it is robust to measurement noise and to clutter induced effects.
It is computationally expensive,  but it works with a reflectivity function that may change sign and it can recover the sign of $\rho(\by)$. The imaging may be potentially improved by adding prior information about $\rho(\by)$ as a penalty term 
in the optimization.
\end{itemize}

\section*{Acknowledgements}
This work is partially supported by the AFOSR grant FA9550-21-1-0166.
\appendix
\section{Random travel time model}
\label{ap:A}
In this appendix we recall from \cite[Chapter 12]{garnier2016passive} and 
\cite{borcea2011enhanced} the geometrical optics ``random travel time" model
of the Green's function $\hat G_\mu(\bz,\bx_n,\om)$ at frequency $\om$ in the bandwidth of the probing signal. Since this bandwidth is much smaller than the central frequency $\om_o$, 
we have $\om/\om_o \approx 1$ and therefore, the wavelength is approximately $\la_o$.

We are interested in points $\bz = (z,\zpar)$ in the support of the reflectivity $\rho$, which are at large range offset of order $L \gg \ell$ from the location $\bx_n = (x_n,L)$ of the platform, for $n = 0, \ldots, N$. The geometrical optics approximation 
is obtained under the high-frequency assumption $\la_o \ll \ell$ and for small fluctuations $\sigma \ll 1$.  The precise scaling assumptions are
\begin{equation}
\sigma^2 \left(\frac{L}{\ell} \right)^3 \ll \frac{\la_o^2}{\sigma^2 \ell L} \ll 1,
\label{eq:A1}
\end{equation}
where the first inequality says that  the fluctuations are weak enough, so the rays are approximately 
straight, the variance of the geometrical spreading amplitude factor of the Green's function is negligible and only the first order in $\sigma$ correction of the travel time matters. However, the second inequality says that $\sigma$ is not too small, so we have significant random fluctuations of the travel time. 

The Green's function approximation is \eqref{eq:A2}, with random travel time correction
\begin{equation}
\tau_\mu(\bz,\bx_n) = \frac{\sigma |\bz-\bx_n|}{2 c} \int_0^1 ds \, \mu \left[ \frac{\bx_n + (\bz-\bx_n)s }{\ell} \right],
\label{eq:A4}
\end{equation}
calculated by integration along the straight ray from $\bx_n$ to $\bz$.   Obviously, it satisfies 
\begin{equation}
\EE\left[\tau_\mu(\bz,\bx_n) \right] = 0.
\label{eq:A5}
\end{equation}
Because $|\bz-\bx_n| = O(L) \gg \ell$, the process $\tau(\by,\bx_n)$  has Gaussian statistics \cite[Lemma 12.1]{garnier2016passive} with variance 
\begin{equation}
\EE\left[ \tau^2_\mu(\bz,\bx_n) \right] = \tau^2 := \frac{\sqrt{2 \pi} \sigma^2 \ell L}{4 c^2},
\label{eq:A6}
\end{equation} 
and covariance
\begin{equation*}
\EE\left[ \tau_\mu(\bz,\bx_n) \tau_\mu (\bz',\bx_{n'}) \right] \approx  \tau^2 \int_0^1 ds \, 
\exp \left\{ - \frac{[(x_n-x_{n'}) s+ (z-z')(1-s)]^2}{2 \ell^2} \right\}.
\end{equation*}
Consequently, the expectation of the random exponential in \eqref{eq:A2} is 
\begin{equation}
\EE\big[ \exp[i \om \tau_\mu(\bz,\bx_n)]\big] = \exp \Big(-\frac{\om^2 \tau^2}{2} \Big) \approx 0,
\label{eq:A8}
\end{equation}
where we used assumption \eqref{eq:A1}. Moreover, after some calculation, 
as shown in \cite[Section 12.1.3]{garnier2016passive}, we obtain 
\begin{align}
&\EE\left[ \exp[i \om \tau_\mu(\bz,\bx_n)-i \om' \tau_\mu(\bz',\bx_{n'})]\right] \approx 
\exp \left[- \frac{(\om-\om')^2 }{8\Omega_d^2} \right.  \nonumber \\
& \qquad \qquad \left. - \frac{ (x_{n'}-x_n)^2 + (x_{n'}-x_n)(z'-z) + (z'-z)^2}{8 \cX_d^2} \right],
\label{eq:A12p}
\end{align}
where we introduced the frequency scale called ``decoherence frequency"
\begin{align}
\Omega_d:= \frac{1}{2\tau} \stackrel{\eqref{eq:A6}}{=} \frac{c}{(2 \pi)^{1/4} \sigma \sqrt{\ell L}} 
\stackrel{\eqref{eq:A1}}{\ll } \frac{c}{(2 \pi)^{1/4} \sqrt{\ell L}} \frac{\sqrt{\ell L}}{\la} = O(\om_o),
\label{eq:A11}
\end{align}
and the length scale called ``decoherence length"
\begin{align}
\cX_d := \frac{\sqrt{3} \ell}{2\om_o \tau} \stackrel{\eqref{eq:A6}}{=} \frac{\sqrt{3} \la_o \sqrt{\ell}}{(2 \pi)^{5/4} \sigma \sqrt{L}} 
\stackrel{\eqref{eq:A1}}{\ll } \frac{\sqrt{3} \la_o \sqrt{\ell}}{(2 \pi)^{5/4}  \sqrt{L}} \frac{\sqrt{\ell L}}{\la_o} = O(\ell).
\label{eq:A12}
\end{align}

Finally, we see that if we make the additional assumption 
\begin{equation}
\sigma \ll \frac{\sqrt{\ell \la_o}}{L},
\label{eq:A13}
\end{equation}
which is comparable to the first inequality in \eqref{eq:A1} if $\ell = O(\sqrt{\la_o L})$ for example,
then we can conclude that if 
\begin{equation}
|z-z'| = O \left(\frac{\la_o L}{\cX_d} \right),
\label{eq:A14}
\end{equation}
we have $|z-z'| \ll \cX_d$ and we can simplify the moment formula \eqref{eq:A12p} as in  \eqref{eq:Mom2}.
Note that the right-hand side in \eqref{eq:A14} is basically the CINT cross-range resolution $H$. In this paper 
we are interested in zooming in a peak of the CINT image,  which is why we assume \eqref{eq:A14}.

\section{Proof that $\cI^{\CINT}$ is nonnegative, real valued}
\label{app:proofICINT}%
By (\ref{eq:2CINT}) the CINT function is given by
$$
\cI^\CINT(\by) =  \sum_{n,n'=0}^N \int_\RR \frac{d \om}{2\pi} \int_\RR \frac{d \om'}{2\pi}  r_n(\by,\om)   \overline{ r_{n'}(\by, \om')} 
 \exp \Big[ -\frac{(x_n-x_{n'})^2}{2 \cX^2} - \frac{(\om-\om')^2}{2 \Omega^2} \Big], 
$$
where $r_n(\by,\om)= \hat R_n(\om) \overline{\hat F_n(\by,\om)} e^{- i \om n T}  $.
As 
\begin{align*}
\exp \Big[ -\frac{(x_n-x_{n'})^2}{2 \cX^2}\Big] &= 
\frac{\sqrt{2}}{\sqrt{\pi} \cX}
\int_\RR dx'' \exp \Big[ -\frac{(x''-x_n)^2}{\cX^2}\Big]\exp\Big[-\frac{(x''-x_{n'})^2}{\cX^2}\Big] ,
\\
\exp \Big[ - \frac{(\om-\om')^2}{2 \Omega^2} \Big]&=
\frac{\sqrt{2}}{\sqrt{\pi} \Omega}
\int_\RR d\om'' 
\exp \Big[ - \frac{(\om''-\om)^2}{\Omega^2} \Big]
\exp \Big[ - \frac{(\om''-\om')^2}{\Omega^2} \Big], 
\end{align*}
we can also write
$$
\cI^\CINT(\by) =\frac{2}{\pi \Omega \cX}  \int_\RR d\om'' \int_\RR dx'' \Big| \sum_{n=0}^N \int_\RR  \frac{d \om}{2\pi}
\exp \Big[ -\frac{(x''-x_n)^2}{\cX^2}  - \frac{(\om''-\om)^2}{\Omega^2} \Big] r_n(\by, \om)\Big|^2 ,
$$
which proves that $\cI^\CINT(\by) \geq 0$.

\section{Estimation of the spectrum of $\cL$}
\label{ap:B}
Here we give the calculations for the results stated in section \ref{sect:spectral}.

\subsection{Proof that  $\cL$ is positive semidefinite}
\label{ap:B1}
To show that 
\begin{equation}
\lb \varphi, \cL \varphi \rb = \int_{\RR} d y \, \overline{\varphi(y)} \cL \varphi(y) \ge 0, \qquad \forall \, 
\varphi \in L^2(\RR),
\label{eq:B1}
\end{equation}
 let us recall from equations \eqref{eq:F9}, \eqref{eq:Mom4} and \eqref{eq:2CINT} 
the calculation of the kernel of $\cL$. With the notation 
$$
r_n (y,\om) := \hat R_n(\om) \overline{\hat F_n(\by,\om)} e^{- i \om n T}  ,
$$
for points $\by = (y,\zpar)$ in the range bin $\cD_{\zpar}$,
we obtain  that 
$$
\mI(y,y') =
\sum_{n,n'=0}^N \int_{\RR} \frac{d \om}{2 \pi}  \int_{\RR} \frac{ d \om'}{2 \pi}
r_n(y,\om) \overline{r_{n'}(y',\om')}
 \exp \left[ -\frac{(x_n-x_{n'})^2}{2 \cX^2} - \frac{(\om-\om')^2}{2 \Omega^2}\right].
$$
Substituting this expression into definition \eqref{eq:S4} of $\cL$, we get 
\begin{align}
\lb \varphi, \cL \varphi \rb = \int_{\RR} dy \, \overline{\varphi(y)} &
 \int_{\RR} dy' \varphi(y')  \sum_{n,n'=0}^N \int_{\RR} \frac{d \om}{2 \pi}   \int_{\RR} \frac{d \om'}{2 \pi} 
r_n(y,\om) \overline{r_{n'}(y',\om')}\nonumber \\
&\times  \exp \left[ -\frac{(x_n-x_{n'})^2}{2 \cX^2} - \frac{(\om-\om')^2}{2 \Omega^2}\right],
\label{eq:B4}
\end{align}
and the goal is to show that this is non-negative.
Let us denote 
$$
r_n(\om) := \int_{\RR} dy \, \overline{\varphi(y)} {r_n(y,\om)} ,
$$
so that \eqref{eq:B4} becomes
\begin{align*}
\lb \varphi, \cL \varphi \rb  = \sum_{n,n'=0}^N
\int_{\RR} \frac{d \om}{2 \pi}   \int_{\RR} \frac{d \om'}{2 \pi}  
\, r_n(\om) \overline{r_{n'}(\om')}  \exp \left[ -\frac{(x_n-x_{n'})^2}{2 \cX^2} - \frac{(\om-\om')^2}{2 \Omega^2}\right].
\end{align*}
From here, the proof that $\lb \varphi, \cL \varphi \rb \geq 0 $ is the same as the proof of the positivity of $\cI^\CINT$ presented in  appendix \ref{app:proofICINT}.
\subsection{Proof of Proposition \ref{prop.1}}
\label{ap:B2}
We begin with an identity, which is verified by direct calculation
\begin{align}
\int_{\RR} d y' \, K_H &\Big(\frac{y+y'}{2}-\frac{z_j+z_{j'}}{2}\Big) K_h \big((y-y')-(z_j-z_{j'})\big) e^{-\frac{(y'-\eta)^2}{2 h H} } = \frac{1}{\sqrt{2 \pi}(H+h/2)} \nonumber \\
&\hspace{0.2in} \times \exp \left\{ - \frac{(z_{j'}-\eta)^2}{2 (H+h/2)^2} - \frac{\Big[y-z_j + (z_{j'}-\eta) 
\Big(\frac{H-h/2}{H+h/2}\Big)\Big]^2}{2 Hh} \right\},
\label{eq:B6}
\end{align}
where $\eta$ is some arbitrary real number. The proof of Proposition \ref{prop.1}  uses this identity and the Hermite polynomials  ${\rm He}_n(\xi)$, for $n \ge 0$. These  polynomials are given by \cite{abramowitz1988handbook}
\begin{equation}
{\rm He}_n(\xi) := (-1)^n e^{{\xi^2}/{2}} \frac{d^n}{d\xi^n} e^{-{\xi^2}/{2}} = \sum_{m=0}^{\lfloor{n/2}\rfloor} 
\theta_{n,n-2m} \xi^{n-2m},
\label{eq:B7}
\end{equation}
with coefficients 
\begin{equation}
\theta_{n,n-2m} =  \frac{(-1)^m n!}{m! (n-2m)!},
\label{eq:B8}
\end{equation}
and they  form an orthogonal basis of $L^2_{w}(\RR)$, the Hilbert space of square integrable functions 
with respect to the weight $w(\xi) := \exp(-\xi^2/2),$
\begin{equation}
\int_{\RR} d\xi \, {\rm He}_n(\xi) {\rm He}_m(\xi) e^{-\xi^2/2} = \sqrt{2 \pi} \, n! \, \delta_{n,m}.
\label{eq:B9}
\end{equation}

\begin{lemma}
\label{lem.1}
Let $\eta$ be an arbitrary real number. We have for all $n \ge 0$, 
\begin{align}
&\int_{\RR} d y' \, K_H \Big(\frac{y+y'}{2}-\frac{z_j+z_{j'}}{2}\Big) K_h \big((y-y')-(z_j-z_{j'})\big) e^{-\frac{(y'-\eta)^2}{2 h H} } {\rm He}_n \Big(\frac{y'-\eta}{\sqrt{Hh}} \Big) \nonumber \\
&= \frac{(H^2+h^2/4)^{n/2}}{\sqrt{2 \pi} (H + h/2)^{n+1}} \exp \left\{- \frac{(z_{j'}-\eta)^2}{2 (H+h/2)^2} - \frac{\Big[y-z_j + (z_{j'}-\eta) 
\Big(\frac{H-h/2}{H+h/2}\Big)\Big]^2}{2 Hh} \right\} \nonumber \\
&\quad \times {\rm He}_n \left[ \frac{(y-z_j)(H-h/2)}{\sqrt{Hh (H^2 + h^2/4)}} + \frac{(z_{j'}-\eta)\sqrt{H^2+h^2/4}}{\sqrt{Hh}(H+h/2)}\right].\label{eq:B10} 
\end{align}
\end{lemma}
\begin{proof} The case $n = 0$ is just the identity \eqref{eq:B6}, because ${\rm He}_0(\xi) = 1$, so let us  consider $n > 1$.
For convenience, denote the left-hand side in \eqref{eq:B10} by $\cF_n$. Substituting the definition \eqref{eq:B7} of the Hermite polynomial into it we get 
\begin{align}
\cF_n = (Hh)^{n/2} \int_{\RR} d y' \, K_H \Big(\frac{y+y'}{2}-\frac{z_j+z_{j'}}{2}\Big) K_h \big((y-y')-(z_j-z_{j'})\big)  
 \nonumber \\
 \times \frac{d^n}{d\eta^n} \exp \Big[-\frac{(y'-\eta)^2}{2  Hh} \Big],
 \label{eq:B11}
 \end{align}
 because
 \[
 \frac{d^n}{d\xi^n} e^{-{\xi^2}/{2}} \Big|_{\xi = \frac{y'-\eta}{\sqrt{Hh}}} = \big(-\sqrt{Hh}\big)^n  \frac{d^n}{d\eta^n} \exp \Big[-\frac{(y'-\eta)^2}{2  Hh} \Big].
 \]
 The integrand is smooth, so we can interchange the integral with the derivative in $\eta$, and then use the identity  \eqref{eq:B6} to rewrite \eqref{eq:B11} as 
 \begin{align*}
 \cF_n = \frac{(Hh)^{n/2}}{\sqrt{2 \pi}(H+h/2)}  \frac{d^n}{d\eta^n}  \exp \left\{ - \frac{(z_{j'}-\eta)^2}{2 (H+h/2)^2} - \frac{\Big[y-z_j + (z_{j'}-\eta) 
\Big(\frac{H-h/2}{H+h/2}\Big)\Big]^2}{2 Hh} \right\}.
\end{align*}
 Grouping all the $\eta$ terms in the exponent in a single square
 \begin{align}
 \frac{(z_{j'}-\eta)^2}{2 (H+h/2)^2} + \frac{\Big[y-z_j + (z_{j'}-\eta) 
\Big(\frac{H-h/2}{H+h/2}\Big)\Big]^2}{2 Hh} = \frac{(y-z_j)^2}{2(H^2 + h^2/4)} \nonumber \\
+\frac{1}{2} \left[ \frac{(z_{j'}-\eta) \sqrt{H^2+h^2/4}}{\sqrt{Hh} (H+h/2)} + 
\frac{(y-z_j)(H-h/2)}{\sqrt{Hh(H^2+h^2/4)}} \right]^2, \label{eq:B13}
\end{align}
we obtain
\begin{align*}
\cF_n = \frac{(Hh)^{n/2}}{\sqrt{2 \pi}(H+h/2)} \exp \left[- \frac{(y-z_j)^2}{2(H^2 + h^2/4)}\right] 
\left[ -\frac{\sqrt{H^2+h^2/4}}{\sqrt{Hh} (H+h/2)} \right]^n \\
\times \frac{d^n}{d\xi^n} e^{-\xi^2/2} \Big|_{\xi = \frac{(z_{j'}-\eta) \sqrt{H^2+h^2/4}}{\sqrt{Hh} (H+h/2)} + 
\frac{(y-z_j)(H-h/2)}{\sqrt{Hh(H^2+h^2/4)}} }.
\end{align*}
The result \eqref{eq:B10} in the lemma follows from this equation, definition \eqref{eq:B7} of the polynomial ${\rm He}_n$ and the 
identity \eqref{eq:B13}.
\end{proof}

To get the eigenfunctions of the operator $\cL_j$ defined in \eqref{eq:S6}, we use Lemma \ref{lem.1} for 
$\eta = z_j = z_{j'}$. We get
\begin{align}
\cL_j e^{-\frac{(y-z_j)^2}{2 h H} } {\rm He}_n \Big(\frac{y-z_j}{\sqrt{Hh}} \Big) &=  \int_{\RR} dy' \, \mI_j(y,y') e^{-\frac{(y'-z_j)^2}{2 h H} } {\rm He}_n \Big(\frac{y'-z_j}{\sqrt{Hh}} \Big)  \nonumber \\
&\hspace{-0.7in}=\frac{C \rho_j^2(H^2+h^2/4)^{n/2}}{\sqrt{2 \pi} (H + h/2)^{n+1}} e^{- \frac{(y-z_j)^2}{ 2 Hh}} 
{\rm He}_n \left[ \frac{(y-z_j)(H-h/2)}{\sqrt{Hh (H^2 + h^2/4)}}\right],
\label{eq:B13p}
\end{align}
which looks almost like an eigenvalue equation, except that the argument of the Hermite polynomial 
in the right-hand side is not $(y-z_h)/\sqrt{Hh}$, but an approximation of it,  since $H \gg h$. Now let us
define the new kernel
\begin{equation}
\mathfrak{K}_j(y,y'):= e^{\frac{(y-z_j)^2}{ 2 Hh}} \mI_j(y,y') e^{-\frac{(y'-z_j)^2}{2 h H} },
\label{eq:B14}
\end{equation}
which is related to $\mI_j(y,y')$ by a similarity transformation, and obtain from 
\eqref{eq:B13p} that
\begin{align}
\int_{\RR} dy' \, \mathfrak{K}_j(y,y') {\rm He}_n \Big(\frac{y'-z_j}{\sqrt{Hh}} \Big) = \frac{C \rho_j^2(H^2+h^2/4)^{n/2}}{\sqrt{2 \pi} (H + h/2)^{n+1}} {\rm He}_n \left[ \frac{(y-z_j)(H-h/2)}{\sqrt{Hh (H^2 + h^2/4)}}\right].
\label{eq:B15}
\end{align}
We use next this result and the completeness of the set  of Hermite polynomials $\{{\rm He}_n(\xi), n \ge 0\}$ to derive the expression of the polynomials $\{p_n(\xi), \, n \ge 0\}$ in Proposition \ref{prop.1}, which satisfy
\begin{equation}
 \int_{\RR} dy' \, \mathfrak{K}_j(y,y') p_n \left(\frac{y'-z_j}{\sqrt{Hh}}\right) = \Lambda_{j,n} p_n \left(\frac{y-z_j}{\sqrt{Hh}}\right), 
 \qquad n \ge 0.
 \label{eq:B16}
 \end{equation}
 
 We see from the definition \eqref{eq:B7} that there is a bijective mapping between the set of Hermite polynomials $\{{\rm He}_n(\xi), n \ge 0\}$ and the set of monomials $\{\xi^n, n \ge 0\}$. Indeed, we have 
 \begin{align}
 \begin{pmatrix}
 {\rm He}_0(\xi) \\
 {\rm He}_2(\xi) \\ 
 {\rm He}_4(\xi) \\ \vdots
 \end{pmatrix} = \begin{pmatrix} 
 \theta_{0,0} & 0 & 0 & 0&\ldots & 0 \\
 \theta_{2,0} & \theta_{2,2} & 0 &0& \ldots & 0 \\
 \theta_{4,0} & \theta_{4,2} & \theta_{4,4} & 0 & \ldots & 0 \\
 && \vdots 
 \end{pmatrix} 
 \begin{pmatrix} 1 \\ \xi^2 \\ \xi^4 \\ \vdots \end{pmatrix}
 ,
 \label{eq:B17}
 \end{align}
 and similarly 
 \begin{align}
 \begin{pmatrix}
 {\rm He}_1(\xi) \\
 {\rm He}_3(\xi) \\ 
 {\rm He}_5(\xi) \\ \vdots
 \end{pmatrix} = \begin{pmatrix} 
 \theta_{1,1} & 0 & 0 & 0&\ldots & 0 \\
 \theta_{3,1} & \theta_{3,3} & 0 &0& \ldots & 0 \\
 \theta_{5,1} & \theta_{5,3} & \theta_{5,5} & 0 & \ldots & 0 \\
 && \vdots 
 \end{pmatrix} 
 \begin{pmatrix} \xi \\ \xi^3 \\ \xi^5 \\ \vdots \end{pmatrix},
 \label{eq:B18}
 \end{align}
 which we write in compact form as 
 \begin{equation}
 {\rm He}_n(\xi) = \sum_{i=0}^n \Theta_{n,i} \xi^i.
 \label{eq:B17-18}
 \end{equation}
Note that $\boldsymbol{\Theta} = \big(\Theta_{n,i}\big)_{n, i \ge 0}$ is lower triangular, with non-zero diagonal entries, so we can invert equation \eqref{eq:B17-18} to get 
\begin{equation}
\xi^i = \sum_{l=0}^i (\boldsymbol{\Theta}^{-1})_{i,l} {\rm He}_l(\xi), \qquad i \ge 0.
\label{eq:B19}
\end{equation}
The polynomials that we seek are of the form 
\begin{equation}
\begin{pmatrix} 
p_0(\xi) \\
p_1(\xi) \\
p_2(\xi) \\ \vdots \end{pmatrix} = \boldsymbol{\Gamma} \begin{pmatrix} 1 \\ \xi \\ \xi^2 \\ \vdots \end{pmatrix},
\label{eq:B20}
\end{equation}
where $\boldsymbol{\Gamma} = (\gamma_{n,i})_{n,i \ge 0}$ is the lower triangular matrix of coefficients that we wish to determine, 
with diagonal $\gamma_{n,n} = 1$. To find these coefficients substitute $p_n$ from \eqref{eq:B20} in \eqref{eq:B16},
\begin{equation*}
\sum_{i=0}^n \gamma_{n,i} \int_{\RR} dy' \, \mI_j(y,y') \left(\frac{y'-z_j}{\sqrt{Hh}} \right)^i 
= \Lambda_{j,n} \sum_{i=0}^n \gamma_{n,i} \left(\frac{y-z_j}{\sqrt{Hh}} \right)^i 
 \end{equation*}
and use \eqref{eq:B19} and \eqref{eq:B15} to get 
\begin{align}
\sum_{i=0}^n \gamma_{n,i} \sum_{l=0}^i (\boldsymbol{\Theta}^{-1})_{i,l} \frac{C \rho_j^2(H^2+h^2/4)^{l/2}}{\sqrt{2 \pi} (H + h/2)^{l+1}}
{\rm He}_l\left[ \frac{(y-z_j)(H-h/2)}{\sqrt{Hh (H^2 + h^2/4)}}\right] = \Lambda_{j,n} \nonumber \\
\times \sum_{i=0}^n \gamma_{n,i} \left(\frac{y-z_j}{\sqrt{Hh}} \right)^i .
\label{eq:B21}
\end{align}
Finally, substituting ${\rm He}_l$ from \eqref{eq:B17-18}, we obtain
\begin{align}
\frac{C \rho_j^2}{\sqrt{2 \pi} (H+h/2)} \sum_{i=0}^n \sum_{l=0}^i \sum_{q=0}^l &\gamma_{n,i} (\boldsymbol{\Theta^{-1}})_{i,l} \left(\frac{\sqrt{H^2+h^2/4}}{H+h/2}\right)^l
\Theta_{l,q} \left(\frac{H-h/2}{\sqrt{H^2+h^2/4}}\right)^q \nonumber \\
&\times \left(\frac{y-z_j}{\sqrt{Hh}}\right)^q = 
\Lambda_{j,n} \sum_{q=0}^n \gamma_{n,q} \left(\frac{y-z_j}{\sqrt{Hh}} \right)^q.
\label{eq:B22}
\end{align}

We can now determine the matrix $\boldsymbol{\Gamma}$ 
and the eigenvalues as follows: Introduce the diagonal matrices 
\begin{align}
\boldsymbol{\cD} :=& \mbox{diag} \left( \left(\frac{\sqrt{H^2+h^2/4}}{H+h/2} \right)^l, ~ l \ge 0 \right), 
\label{eq:B24}
\\
{\bf D}:= & \mbox{diag} \left( \left(\frac{H-h/2}{\sqrt{H^2+h^2/4}}\right)^q, ~ q \ge 0 \right),
\label{eq:B25}
\end{align}
and let also 
\begin{equation}
\boldsymbol{\tilde{\Lambda}_j} :=  \mbox{diag} \left( \tilde{\Lambda}_{j,n} := \frac{\Lambda_{j,n} \sqrt{2 \pi}(H+h/2)}{C \rho_j^2}, ~n \ge 0\right).
\end{equation}
Then, since the monomials are linearly independent, we obtain from \eqref{eq:B22} that 
\begin{equation}
\boldsymbol{\Gamma} \boldsymbol{\Theta}^{-1} \boldsymbol{\cD} \boldsymbol{\Theta} {\bf D} = \boldsymbol{\tilde \Lambda}_j \boldsymbol{\Gamma}.
\label{eq:B27}
\end{equation}
Let us look at the diagonal part of the matrix equality \eqref{eq:B27}: The left-hand side gives
\begin{align}
\left(\boldsymbol{\Gamma} \boldsymbol{\Theta}^{-1} \boldsymbol{\cD} \boldsymbol{\Theta} {\bf D}  \right)_{n,n} &= \sum_{i=0}^n \gamma_{n,i} \left(\boldsymbol{\Theta}^{-1} \boldsymbol{\cD} \boldsymbol{\Theta} {\bf D} \right)_{i,n} = \gamma_{n,n} \left(\boldsymbol{\Theta}^{-1} \boldsymbol{\cD} \boldsymbol{\Theta} {\bf D} \right)_{n,n} \nonumber \\
&= \left(\boldsymbol{\Theta}^{-1} \boldsymbol{\cD} \boldsymbol{\Theta} {\bf D} \right)_{n,n},  
\label{eq:B28}
\end{align}
where the second equality is because $\boldsymbol{\Theta}^{-1} \boldsymbol{\cD} \boldsymbol{\Theta} {\bf D} $ is lower triangular, and the last equality is 
because the diagonal entries of $\boldsymbol{\Gamma}$ are $\gamma_{n,n} = 1.$ The right-hand side in \eqref{eq:B27} gives
\begin{equation}
\big(\boldsymbol{\tilde \Lambda_j \Gamma} \big)_{n,n} = \tilde{\Lambda}_{j,n} \gamma_{n,n} = \tilde{\Lambda}_{j,n} .
\label{eq:B29}
\end{equation}
Therefore, 
\begin{align}
\tilde{\Lambda}_{j,n} = \frac{\Lambda_{j,n} \sqrt{2 \pi}(H+h/2)}{C \rho_j^2} &= \left(\boldsymbol{\Theta}^{-1} \boldsymbol{\cD} \boldsymbol{\Theta} {\bf D} \right)_{n,n} = 
\Theta_{n,n}^{-1} \cD_{n,n} \Theta_{n,n} D_{n,n} \nonumber \\
&= \left(\frac{H-h/2}{H+h/2}\right)^n,
\label{eq:B30}
\end{align}
where we used that $\boldsymbol{\Theta}$ is lower triangular and definitions \eqref{eq:B24}--\eqref{eq:B25}. This verifies the 
expressions \eqref{eq:S7} of the eigenvalues in Proposition \ref{prop.1}.

Equating the off-diagonal entries in \eqref{eq:B27} we get
\begin{align*}
\left(\boldsymbol{\Gamma} \boldsymbol{\Theta}^{-1} \boldsymbol{\cD} \boldsymbol{\Theta} {\bf D} \right)_{n,l} &= \sum_{i=0}^n \gamma_{n,i} \left(\boldsymbol{\Theta}^{-1} \boldsymbol{\cD} \boldsymbol{\Theta} {\bf D} \right)_{i,l} =  \left(\boldsymbol{\tilde \Lambda_j \Gamma}\right)_{n,l} = \tilde\Lambda_{j,n}\gamma_{n,l}.
\end{align*}
Let us consider the case $l = n-1$ and use that $\boldsymbol{\Theta}^{-1} \boldsymbol{\cD} \boldsymbol{\Theta} {\bf D}$ is lower triangular, to conclude that only the terms with $i \in \{n-1,n\}$ contribute to the sum. Using again that $\gamma_{n,n}= 1$, we get 
\begin{align*}
\left(\boldsymbol{\Theta}^{-1} \boldsymbol{\cD} \boldsymbol{\Theta} {\bf D} \right)_{n,n-1} + \gamma_{n,n-1} \left(\boldsymbol{\Theta}^{-1} \boldsymbol{\cD} \boldsymbol{\Theta} {\bf D}  \right)_{n-1,n-1} = \tilde \Lambda_{j,n}\gamma_{n,n-1},
\end{align*}
where we note from \eqref{eq:B30} that 
$
 \left(\boldsymbol{\Theta}^{-1} \boldsymbol{\cD} \boldsymbol{\Theta} {\bf D}  \right)_{n-1,n-1} = \tilde{\Lambda}_{j,n-1}
$.
Therefore, we get 
\begin{equation}
\gamma_{n,n-1} = \frac{\left(\boldsymbol{\Theta}^{-1} \boldsymbol{\cD} \boldsymbol{\Theta} {\bf D}  \right)_{n,n-1} }{\tilde{\Lambda}_{j,n}-\tilde{\Lambda}_{j,n-1}}.
\label{eq:B31}
\end{equation}
Now we can proceed recursively, and obtain using similar calculations that 
\begin{equation}
\gamma_{n,l} = \frac{\sum_{q=l+1}^n \gamma_{n,q} \left(\boldsymbol{\Theta}^{-1} \boldsymbol{\cD} \boldsymbol{\Theta} {\bf D}  \right)_{q,l} }{\tilde{\Lambda}_{j,n}-\tilde{\Lambda}_{j,l}}, \qquad l = n-2, n-3, \ldots, 0.
\label{eq:B32}
\end{equation}
 \subsection{Proof of Proposition \ref{prop.2}}
\label{ap:B3}
Recalling the definitions \eqref{eq:S3} and \eqref{eq:S5} of the integral operators $\cL$ and $\cL_j$, we see that we can write
\begin{equation}
\cL = \sum_{j=1}^M \cL_j + \sum_{j,j' = 1}^M (1-\delta_{j,j'})\cL_{j,j'},
\label{eq:B33}
\end{equation}
where $\cL_{j,j'} : L^2(\RR) \mapsto L^2(\RR)$ are the integral operators 
\begin{equation}
\cL_{j,j'} \varphi(y) = \int_{\RR} dy' \, \mI_{j,j'}(y,y') \varphi(y'), \qquad \forall \, \varphi \in L^2(\RR),
\label{eq:B34}
\end{equation}
with kernels
\begin{equation}
\mI_{j,j'}(y,y') = C \rho_j \rho_{j'} K_H\left(\frac{y+y'}{2}-\frac{z_j+z_{j'}}{2}\right) K_h\left((y-y')-(z_j-z_{j'})\right), 
\label{eq:B35}
\end{equation}
for $j,j' \in \{1,\ldots, M\}$, with $j \ne j'$.

Now let us assume that the separation condition  \eqref{eq:S11} holds. From Lemma \ref{lem.1}, the expression 
\eqref{eq:S8} of the eigenfunctions of $\cL_j$ and the definition of the polynomials $p_n(\xi)$ in terms of the Hermite polynomials given above we obtain that 
\begin{align}
\int_{\RR} d y' \, K_H\left(\frac{y+y'}{2}-\frac{z_j+z_{j'}}{2}\right) K_h\left((y-y')-(z_j-z_{j'})\right) V_{l,n}(y') \nonumber \\
=
\left\{ \begin{array}{ll}
\frac{\left(\frac{H-h/2}{H+h/2}\right)^n}{\sqrt{2 \pi}(H+h/2)} V_{j,n}(y), \quad &\mbox{if} ~ l = j' \\\\
O\left(e^{-9 \zeta^2/2} \right) \ll 1, \quad &\mbox{otherwise}.
\end{array} \right.
\label{eq:B36}
\end{align}
Using this result, equation \eqref{eq:B33} and the linearity of the operators, we get 
\begin{align}
\left[\cL \sum_{j=1}^M \rho_j V_{j,n}\right](y) &=\sum_{j=1}^M \rho_j \left[ \sum_{l=1}^M \cL_l V_{j,n}(y) + 
\sum_{l,l' = 1}^M (1-\delta_{l,l'})\cL_{l',l}V_{j,n}(y)\right] \nonumber \\
&\hspace{-0.7in}=  \sum_{l=1}^M \rho_l \cL_l V_{l,n}(y) + \sum_{l,l'=1}^M (1-\delta_{l,l'})\rho_l \cL_{l',l} V_{l',n}(y) + O\left(e^{-9 \zeta^2/2}\right).
\end{align}
Now we use Proposition \ref{prop.1} and equations \eqref{eq:B34}--\eqref{eq:B36} to obtain 
\begin{align}
\left[\cL \sum_{j=1}^M \rho_j V_{j,n}\right](y) \approx \sum_{l'=1}^M \left[\sum_{l=1}^M \frac{C \rho_l^2\left(\frac{H-h/2}{H+h/2}\right)^n}{\sqrt{2 \pi} (H+h/2)}\right] \rho_{l'} 
V_{l',n}(y) + O\left(e^{-9 \zeta^2/2}\right).
\end{align}
This is the statement in Proposition \ref{prop.2}, where we recognize that the term within the square brackets defines the eigenvalue $\Lambda_n$ given in \eqref{eq:S12} and the eigenfunction $V_n(y)$ is as  in \eqref{eq:S13}.

To complete the proof, it remains to show that $\{V_n(y), ~ n \ge 0\}$ are the only eigenfunctions of $\cL$. We use 
the analytic perturbation theory for self-adjoint linear operators \cite{kato2013perturbation}, and proceed recursively with respect to $M$:

When  $M = 1$, the operator is  $\cL^{(1)} = \cL_1$, and its spectrum is given in Proposition \ref{prop.1} for $j = 1$, with the eigenfunctions $\{V_{1,n}(y), ~n \ge 0 \}$.  Note that since the polynomials are a complete set,  these eigenfunctions form an orthogonal basis of $L^2(\RR)$. 

When $M = 2$, the operator is $\cL^{(2)} = \cL^{(1)} + \cL_2 + \cL_{1,2} + \cL_{2,1}$ and depends analytically on $\rho_2$. 
Since there is no other eigenfunction for $\rho_2 = 0$, there is no other eigenfunction that emerges for $\rho_2 \ne 0$.  
Proceeding this way, we get at step $M-1$ the recursive hypothesis that  the eigenfunctions of 
the operator $\cL^{(M-1)}$ are as in 
Proposition \ref{prop.2}, for $M$ replaced by $M-1$. Then, 
\[
\cL^{(M)} = \cL^{(M-1)} + \cL_M +  \sum_{j = 1}^M (1-\delta_{j,M}) 
\big(\cL_{j,M} + \cL_{M,j} \big)\] depends analytically on $\rho_M$, and the analytic perturbation theory \cite{kato2013perturbation} gives that all the eigenfunctions are $\{V_n(y), ~ n \ge 0\}$, as stated in Proposition \ref{prop.2}.

\bibliographystyle{siam} \bibliography{NEW_CINT.bib}

\end{document}